\documentclass{aastex63}
\newcommand{\MyFigA}{\ref{MyFigA}}
\newcommand{\MyFigB}{\ref{MyFigB}}
\newcommand{\MyFigC}{\ref{MyFigC}}
\newcommand{\MyTabA}{\ref{MyTabA}}
\newcommand{\MyTabB}{\ref{MyTabB}}
\newcommand{\MyTabC}{\ref{MyTabC}}
\begin{document}
\title{Jets in a Gamma-Ray Burst During its Prompt Emission: Evolution of Lorentz Factor}
\correspondingauthor{Da-Bin Lin, Rui-Jing Lu, En-Wei Liang}
\email{lindabin@gxu.edu.cn, luruijing@gxu.edu.cn, lew@gxu.edu.cn}
\email{}
\author{Jing Li}
\affil{Laboratory for Relativistic Astrophysics, Department of Physics, Guangxi University, Nanning 530004, China}
\author{Da-Bin Lin}
\affil{Laboratory for Relativistic Astrophysics, Department of Physics, Guangxi University, Nanning 530004, China}
\author{Rui-Jing Lu}
\affil{Laboratory for Relativistic Astrophysics, Department of Physics, Guangxi University, Nanning 530004, China}
\author{Yun Wang}
\affiliation{Key Laboratory of Dark Matter and Space Astronomy, Purple Mountain Observatory, Chinese Academy of Sciences, Nanjing 210034, China}
\affiliation{School of Astronomy and Space Science, University of Science and Technology of China, Hefei, Anhui 230026, China}
\author{Lu-Yao Jiang}
\affiliation{Key Laboratory of Dark Matter and Space Astronomy, Purple Mountain Observatory, Chinese Academy of Sciences, Nanjing 210034, China}
\affiliation{School of Astronomy and Space Science, University of Science and Technology of China, Hefei, Anhui 230026, China}
\author{Shen-Shi Du}
\affiliation{School of Physics and Technology, Wuhan University, Wuhan, Hubei 430072, China;}
\author{Wen-Qiang Liang}
\affil{Laboratory for Relativistic Astrophysics, Department of Physics, Guangxi University, Nanning 530004, China}
\author{Xiang-Gao Wang}
\affil{Laboratory for Relativistic Astrophysics, Department of Physics, Guangxi University, Nanning 530004, China}
\author{En-Wei Liang}
\affil{Laboratory for Relativistic Astrophysics, Department of Physics, Guangxi University, Nanning 530004, China}
\begin{abstract}
Knowledge about the Lorentz factor and its evolution of relativistic jets in gamma-ray bursts (GRBs)
is crucial to understand their physics.
An exact value of bulk Lorentz factor can be estimated based on a high-energy spectral cutoff,
which may appear in GRBs' prompt emission owing to the absorption of photon-photon pair production.
In this work, we focus on the investigation of the bulk Lorentz factor evolution of jets in an individual burst.
Based on \textsl{Fermi} observations,
we search for the bursts with multiple $\gamma$-ray pulses
characterized by a high-energy spectral cutoff,
and nine GRBs are obtained.
Together with the estimation of the pulse duration and radiation spectrum,
the Lorentz factor of jets corresponding to different pulses in an individual GRB are estimated.
It is shown that the Lorentz factor of jets in an individual GRB fluctuates
within a certain range and without a general trend in these nine GRBs.
In addition, the Lorentz factors of the jets in GRBs~130821A, 160509A and 160625B seem to increase with time.
We also study the relations among $L_{\rm iso }$, $E_{\rm p,z}$, and $\Gamma$ for the pulses in our sample,
which is found to be consistent with that found in previous works.
\end{abstract}

\section{Introduction}\label{Sec:Intro}
Gamma-ray bursts(GRBs) are among the most powerful explosions in the universe
(e.g., \citealp{Piran_T-2004-RvMP.76.1143P, Zhang_B-2007-ChJAA.7.1Z}).
It was early realized on that the phenomena of GRBs are associated with an relativistic jet
(\citealp{Krolik_JH-1991-Pier_EA-ApJ.373.277K}; \citealp{Fenimore_EE-1993-Epstein_RI-A&AS.97.59F}; \citealp{Woods_E-1995-Loeb_A-ApJ.453.583W}; \citealp{Baring_MG-1997-Harding_AK-ApJ.481L.85B}).
However, the physical origin for the prompt emission of GRBs is still controversial.
Several scenarios have been proposed, e.g., the internal shock formed in an erratic relativistic fireball (\citealp{Rees_MJ-1994-Meszaros_P-ApJ.430L.93R}; \citealp{Paczynski_B-1994-Xu_G-ApJ.427.708P}),
the photosphere of the fireball (\citealp{Thompson_C-1994-MNRAS.270.480T}; \citealp{Ghisellini_G-1999-Celotti_A-ApJ.511L.93G}; \citealp{Peer_A-2006-Meszaros_P-ApJ.642.995P}; \citealp{Thompson_C-2007-Meszaros_P-ApJ.666.1012T}; \citealp{Giannios_D_-2008-A&A.480.305G}; \citealp{Lazzati_D-2010-Begelman_MC-ApJ.725.1137L};
\citealp{Mizuta_A-2011-Nagataki_S-ApJ.732.26M}; \citealp{Lazzati_D-2013-Morsony_BJ-ApJ.765.103L}; \citealp{Ruffini_R-2013-Siutsou_IA-ApJ.772.11R}),
the discharged magnetic energy of a Poynting-flux-dominated jet (e.g., \citealp{Spruit_HC-2001-Daigne_F-A&A.369.694S,Drenkhahn_G-2002-Spruit_HC-A&A.391.1141D,Giannios_D_-2008-A&A.480.305G,
Zhang_B-2011-Yan_H-ApJ.726.90Z,McKinney_JC-2012-Uzdensky_DA-MNRAS.419.573M,Kumar_P-2015-Crumley_Patrick_-MNRAS.453.1820K,
Sironi_L-2016-Giannios_D-MNRAS.462.48S,Beniamini_P-2016-Granot_J-MNRAS.459.3635B,Granot_J-2016-ApJ.816L.20G}
),
the external shock formed during the deceleration of the jet (\citealp{Burgess_JM-2016-Begue_D-ApJ.822.63B,Huang_LY-2018-Wang_XG-ApJ.859.163H}),
or a hadronic scenario that the synchrotron radiation from a population of relativistic protons
gives rise to the observed prompt emission (\citealp{Oganesyan_G-2019-Nava_L-A&A.628A.59O,Ghisellini_G-2020-Ghirlanda_G-A&A.636A.82G,Florou_I-2021-Petropoulou_M-MNRAS.505.1367F}).
A GRB jet is either Poynting-flux-dominated or matter-dominated,
and suffered from acceleration at the expense of dissipating its magnetic or thermal energy.
After the initial acceleration, the jet may enter the coasting phase,
maintaining a constant Lorentz factor
until it sweeps up a certain mass from the circum-burst medium to reach the deceleration radius.
Beyond the deceleration radius,
the Lorentz factor of the GRB jet decreases significantly,
characterized by a power-law decay with respect to the observation time.
Corresponding to different scenario for the prompt emission,
the Lorentz factor evolution of a jet or jets in an individual GRB may be very different.
In the internal shock scenario for the prompt emission,
the Lorentz factor of jets corresponding to different pulse in an individual GRB
reveals the fluctuation of the central engine of GRB.
Accompanying the discharged of the magnetic energy,
different pulses of the prompt emission can be produced and
thus the Lorentz factor corresponding to different pulses may reveal the fluctuation in the emission region.

Since Lorentz factor varies depending on the phase of GRBs,
the Lorentz factor $\Gamma$ is a critical parameter in understanding the physics of GRBs.
Several methods for estimating the Lorentz factor of GRB jets have been proposed in the literatures:
(1) The widely used method is to employ the onset bump of the afterglows,
which signals the deceleration of the GRB jets.
In this scenario, the bulk Lorentz factor $\Gamma_0$ of the jets after producing the prompt emission
is related to the peak time of the onset bump
(e.g., \citealp{Meszaros_P-1993-Rees_MJ-ApJ.405.278M, Sari_R-1999-Piran_T-ApJ.520.641S, Zhang_B-2003-Kobayashi_S-ApJ.595.950Z, Molinari_E-2007-Vergani_SD-A&A.469L.13M, Lu_J-2012-Zou_YC-ApJ.751.49L}).
It should be noted that the value of $\Gamma_0$ corresponds to the mean value of jets' Lorentz factor.
(2) The ``compactness problem'' of GRBs is also used to infer the Lorentz factor of jets.
It is early expected that the high-energy photons would be absorbed to produce $e^+e^-$ pairs
if the Lorentz factor of GRB jet is not large enough.
Then, the observations of higher energy photons would give stringent constraints on the Lorentz factor of jet.
In general, a high-energy spectral cutoff would appear in the observed radiation spectrum owing to
the abortion of high-energy photons.
Then, the high-energy spectral cutoff is also used to estimate the Lorentz factor of the GRB jet
(\citealp{Krolik_JH-1991-Pier_EA-ApJ.373.277K, Fenimore_EE-1993-Epstein_RI-A&AS.97.59F, Woods_E-1995-Loeb_A-ApJ.453.583W,Baring_MG-1997-Harding_AK-ApJ.491.663B};
\citealp{Lithwick_Y-2001-Sari_R-ApJ.555.540L}; \citealp{Baring_MG-2006-ApJ.650.1004B}; \citealp{Ackermann_M-2011-Ajello_M-ApJ.729.114A,Ackermann_M-2013-Ajello_M-ApJS.209.11A,Tang_QW-2015-Peng_FK-ApJ.806.194T}; \citealp{Lin_DB-2019-Lu_RJ-ApJ.883.187L}).
(3) Theoretically, the photosphere of a GRB jet is related to the Lorentz factor of the jet.
Thus, measuring the temperature and flux of the thermal components in the early stages of the fireball (before the break)
can limit the outflow bulk Lorentz factor of fireball model parameters
(e.g., \citealp{Peer_A-2007-Ryde_F-ApJ.664L.1P, Zou_YC-2015-Cheng_KS-ApJ.800L.23Z,Deng_LT-2022-Lin_DB-arXiv220508737D,Wang_Y-2022-Zheng_TC-arXiv220508427W}). However, only several bursts are believed to be confidently identified with the thermal emission in the prompt emission.
(4) For most of GRBs, the observations reveal that the emission from the external shock is not brighter than the prompt emission. Then, an upper limit of $\Gamma_{\rm 0}$ for different environment
can be derived according to the standard internal-external shock mode (e.g., \citealp{Zou_YC-2010-Piran_T-MNRAS.402.1854Z}).
Of course, some correlations about the Lorentz factor have been studied successively and used to estimate the Lorentz factor of GRB's jet (e.g., \citealp{Mu_HJ-2016-Lin_DB-ApJ.831.111M}).
\cite{Liang_EW-2010-Yi_SX-ApJ.725.2209L} have discovered a correlation $\Gamma_{\rm 0}\propto E_{\rm \gamma,iso,52}^{0.25}$.
By including more GRBs with update data,
\cite{Lu_J-2012-Zou_YC-ApJ.751.49L} discovered an even tighter correlation
$\Gamma_{\rm 0} \simeq 249 L_{\rm \gamma,iso,52}^{0.3}$.
Moreover, \cite{Liang_EW-2015-Lin_TT-ApJ.813.116L} proposed a tight correlation
among $L_{\rm iso }$, $E_{\rm p,z}$, and $\Gamma_0$ based on a multiple regression analysis,
which take the form of $L_{\rm iso} \propto E_{\rm p,z}^{1.34\pm 0.14} \Gamma_0^{1.32\pm 0.19}$
or $E_{\rm p,z} \propto L_{\rm iso}^{0.55\pm 0.06}\Gamma_0^{-0.5\pm 0.17}$.
These empirical correlations can provide a rough estimation about the Lorentz factor of GRBs' Jet.

The above literatures are mainly focused on the Lorentz factor of GRBs' jet in a certain phase.
The evolution of the Lorentz factor in an individual burst has not been studied extensively (e.g., \citealp{Lin_DB-2019-Lu_RJ-ApJ.883.187L}), especially for the prompt emission phase.
In this paper, we study the evolution of jets' Lorentz factor in an individual burst,
assuming that the high-energy spectral cutoff is induced by
the absorption of two-photon pair production ($\gamma\gamma \leftrightarrow e^+e^-$).
This paper is organized as follows.
In Section~\ref{Sec:Sample and Data}, we present the sample selection and data reduction for bursts with high-energy spectral cutoff.
In Section~\ref{Sec:Light-curve and Radiation Spectrum}, we present the light-curve fitting and spectral analysis
about the pulses of GRBs in our sample.
In Sections~\ref{Sec:Lorentz Factor}, the Lorentz factor of jet corresponding to different pulses is estimated.
Correspondingly, the evolution of Lorentz factors in the process of burst appears.
With the total of nine GRBs having measurement of the Lorentz factors and of which 34 pulses have the characteristics of high-energy spectral cutoff, we further test the correlations among $L_{\rm iso }$, $E_{\rm p,z}$, and $\Gamma$.
The value of $\Gamma_{0,\rm ps}$ is estimated for different GRBs and compare with the initial Lorentz factor $\Gamma _{0,\rm es}$ estimated based on the external-shock afterglow, which are also presented in this section.
In Sections~\ref{Sec:Conclusions}, we present the conclusions.

\section{Sample selection and Data Reduction}\label{Sec:Sample and Data}
The \textsl{Fermi} satellite includes two instruments, Gamma-ray Burst Monitor (GBM, \citealp{Meegan_C-2009-Lichti_G-ApJ.702.791M}) and Large Area Telescope
(LAT, \citealp{Atwood_WB-2009-Abdo_AA-ApJ.697.1071A}) instruments,
and provide unprecedented spectral coverage for seven orders of magnitude in photon energy (from $\sim8$~keV to $\sim300$~GeV). GBM has 12 sodium iodide (NaI) scintillation detectors covering the 8~keV-1~MeV energy band,
and two bismuth germanate (BGO) scintillation detectors that are sensitive to the 200~keV-40~MeV energy band
(\citealp{Meegan_C-2009-Lichti_G-ApJ.702.791M}).
The energy coverage of LAT is 20~MeV-300~GeV.
Since the launch of \textsl{Fermi} satellite, more than 2620 GRBs were detected by the GBM.
Over one hundred bursts have been co-detected by the GBM and LAT,
but only 77~bursts were recorded in \textsl{Fermi} LAT Low-Energy Events catalog (LLE).
We search for the bursts with high-energy spectral cutoff due to two-photon pair production,
which generally appears in the high-energy range $\gtrsim 10$~MeV.
Then, the GRBs simultaneously observed by GBM and LLE would be the candidate in our search.
The python source package \emph{gtBurst} is used to extract the light-curves
and source spectra of GBM and LAT from their Time-Tagged Events (TTE) data, respectively.
The joint spectral fittings for different pulses are performed.
We select the burst of which at least two pulses are presented with a high-energy spectral cutoff in their radiation spectrum.
There are nine GRBs (GRBs~090323, 090926A, 100724B, 120226A, 130821A, 160509A, 160625B, 170405A, and 180720B)
in our sample, of which six have redshift measurement and three have no redshift measurement.
In our analysis, the redshift of $z=1.0$ is set for bursts without redshift measurement.
GRB~130427A and GRB~190114C are among most powerful bursts with the highest-energy photons around 94~GeV and 22.9~GeV respectively based on Fermi-LAT detection
(e.g., \citealp{Zhu_S-2013-Racusin_J-GCN.14471.1Z,Ackermann_M-2014-Ajello_M-Sci.343.42A,Kocevski_D-2019-Omodei_N-GCN.23709.1K,MAGICCollaboration_MAGICC-2019-Acciari_VA-Natur.575.455M}). However, GRB~130427A is very difficult to distinguish pulses
due to the multi-peaked structure in the period of [6.0, 11.5]~s after the Fermi trigger.
More importantly, the time-resolved spectra of this burst could not
be fitted with Band+cutoff model (see Equation~(\ref{Eq:BandCutoff})).
The comprehensive spectral analysis about this burst can be found in \cite{Ackermann_M-2014-Ajello_M-Sci.343.42A}.
The comprehensive spectral analysis of GRB~190114C can be found in \cite{Ursi_A-2020-Tavani_M-ApJ.904.133U} and \cite{Ajello_M-2020-Arimoto_M-ApJ.890.9A}.
We also performed the spectral analysis of the pulses in this burst.
It is found that the radiation spectrum of the pulses could be fitted with Band,
Band+PL, or Band+CPL rather than Band+cutoff,
where PL and CPL represent the power-law and cutoff power-law spectral models, respectively.
In addition, the radiation spectrum in some pulses of this burst seems to be complicated.
Therefore, GRB~130427A and GRB~190114C are not discussed in this paper.

We download data from the FSSC (Fermi Science Support Center)\footnote{https://fermi.gsfc.nasa.gov/ssc/data/access/}.
The light-curves are extracted with the standard HEASOFT tools XSPEC command \emph{gtbin}.
For LAT light-curves, we performed a standard photon selections based on
energy, Region Of Interest (ROI), time, and zenith, which removing the effects of the earth's limb.
During the spectral analysis, radiation spectra are extracted with the \emph{gtBurst}\footnote{https://fermi.gsfc.nasa.gov/ssc/data/analysis/scitools/gtburst.html},
and TTE data from the brightest NaI and BGO detector and public LLE
are used, if the LAT data were available, we also included them in the joint spectral analysis.
For the standard LAT data, instrument response function P8R3$\_$SOURCE$\_$V2 are used.
We adopt the photon above 100~MeV in a ROI of 12 degree,
and exclude the events with zenith angles $>$100$^\circ$ in order to avoid
contribution of Earth-limb gamma rays. Then joint spectral fitting of GBM and LAT data is performed with XSPEC,
which judges the goodness of fit using the ``Poisson-Gauss'' fit statistic (i.e., PGSTAT).

\section{Light-curve and Radiation Spectrum Fittings for Pulses in our Sample}\label{Sec:Light-curve and Radiation Spectrum}
The light-curves of our selected bursts are shown in Figure~{\MyFigA} with black lines.
In this paper, we study the evolution of jets' Lorentz factor corresponding to different pulses in an individual GRB.
Then, we first decompose the light-curves into different pulses.
To identify a pulse, we employ an empirical pulse model
(\citealp{Kocevski_D-2003-Ryde_F-ApJ.596.389K,Lu_RJ-2018-Liang_YF-ApJ.865.153L}), i.e.,
\begin{equation}\label{KRL}
F_{n}(t)=F_{\rm p}\left(\frac{t-t_0}{t_{\rm p}-t_0}\right)^r\left[\frac{d}{d+r}+\frac{r}{d+r}\left(\frac{t-t_0}{t_{\rm p}-t_0}\right)^{r+1}\right]^{-\frac{r+d}{r+1}},
\end{equation}
where $t_0$ measures the offset of the $n$th pulse zero time relative to the GBM trigger time (i.e., $F_n(t)=0$ if $t<t_0$), $t_{\rm p}$ is the time of the peak flux ($F_{\rm p}$),
and $r$ and $d$ are the power-law rising and decaying indices, respectively.
The values of $t_0$, $t_{\rm p}$, $F_{\rm p}$, $r$, and $d$ are generally different for different pulses.
The labels of pulses in a burst are in chronological order for their peak time.
If the number of pulses in a burst is $N$, the total light-curves of this burst can be described as
\begin{equation}\label{Eq:LC_Fitting}
{F_{{\rm{tot}}}}(t) = \sum\limits_{n = 1}^N {{F_n}(t)}  + {F_{\rm{0}}}.
\end{equation}
The light-curve fitting is processed based on the light-curve of counts in the energy channel 8-1000~keV
and the results are shown in Figure~{\MyFigA} with red lines.
The light-curve fitting result for each single pulse in an individual burst of our sample is also reported in Table~{\MyTabA}.
With the light-curve fitting results, the full width at the half maximum(FWHM) $\delta t$ of a pluse can be estimated
and is used as the duration of the corresponding pulse in our estimation of Lorentz factor.
The value of $\delta t$ for different pulses is our main focus in the light-curve fitting
and is reported in the eighth column of Table~{\MyTabA}.

After identifying a pulse,
we perform the joint spectral analysis in order to estimate the radiation spectrum and the high-energy cutoff energy (or its lower limit).
Generally, the radiation spectrum of a GRB can be well fitted with a smoothly connected broken power law,
which is known as the Band function (\citealp{Band_D-1993-Matteson_J-ApJ.413.281B}).
If there is absorption by the two-photon pair production for high-energy photons,
a high-energy spectral cutoff would appear in the radiation spectrum.
Then, we adopt the Band+cutoff spectral model, i.e.,
\begin{equation}\label{Eq:BandCutoff}
{N_E} = N_0\left\{ {\begin{array}{*{20}{c}}
{\left( \frac{E}{1\rm keV}\right)^\alpha\exp \left( { - \frac{E}{{{E_0}}}} \right),}&{E < \frac{{{E_0}{E_{\rm c}}}}{{{E_{\rm c}} - {E_0}}}\left( {\alpha  - \beta } \right),}\\
{{K_2}{\left( \frac{E}{1\rm keV}\right)^\beta }\exp \left( { - \frac{E}{E_{\rm c}}} \right),}&{E > \frac{{{E_0}{E_{\rm c}}}}{{{E_{\rm c}} - {E_0}}}\left( {\alpha  - \beta } \right),}
\end{array}} \right.
\end{equation}
with
\[{K_2} = {\left[ {\frac{{{E_0}{E_{\rm c}}}}{{{E_{\rm c}} - {E_0}}}\left( {\alpha  - \beta } \right)} \right]^{\alpha  - \beta }}\exp \left( {\beta  - \alpha } \right),\]
in our spectral analysis.
Equation~(\ref{Eq:BandCutoff}) would be reduced to the Band function if $E_{\rm c}$ is significantly high compared with the maximum photon energy observed by the \emph{Fermi} satellite.
For the pulses in our sample, we will use Band or Band+cutoff spectral model to fit the radiation spectrum,
where Equation~(\ref{Eq:BandCutoff}) with $E_{\rm c}=10^{12}$~keV is used as the Band spectral model in our fittings.
As an example, Figure~{\MyFigB} presents the spectral fitting results of the pulses in GRB~160509A.
One can find that the high-energy spectral cutoff is obvious in some pulses of this burst.
All of our spectral fitting results are reported in Table~{\MyTabB}.
It can be found that three GRBs, i.e., GRBs~090926A, 120226A, and 170405A, in our sample
have two pulses detected with high-energy spectral cutoff,
two GRBs, i.e., GRBs~090323 and 160625B,
have three pulses detected with high-energy spectral cutoff,
and four GRBs (i.e., GRBs~100724B, 130821A, 160509A and 180720B)
have more pulses detected with high-energy spectral cutoff.
The high-energy spectral cutoff ranges from 19.21 to 702.56~MeV for the pulses in our sample.

\section{Lorentz Factor Estimation and Evolution in a Burst}\label{Sec:Lorentz Factor}
\subsection{Lorentz Factor Estimation and Correlations}
\emph{Method to estimate Lorentz factor}.$\;\;$
In the scenario that two-photon pair production is responsible to the high-energy spectral cutoff,
one can estimate the Lorentz Factor of jet for the corresponding pulse by taking $\tau_{\gamma\gamma}(E_c)=1$.
The photoabsorption optical depth $\tau_{\gamma\gamma}$ of photons with energy $E_c$ from low-energy photons emitted cospatially in the jet shell is given by
(\citealp{Abdo_AA-2009-Ackermann_M-Sci.323.1688A}),
\begin{equation}
{\tau_{\gamma \gamma}}({E_{\rm{c}}}) = {\sigma _T}{(\frac{{{d_L}}}{R})^2}\frac{{{E_{\rm ch}}f({E_{\rm ch}})}}{{{{(1 + z)}^{2\beta  + 4}}}}{(\frac{{{E_{\rm{c}}}{E_{\rm ch}}}}{{{\Gamma ^2}m_e^2{c^4}}})^{ - \beta  - 1}}F(\beta ),
\end{equation}
where $\sigma _T$ is the Thomson scattering cross-section,
$d_L$ is the distance of the burst relative to the observer,
$R$ is the distance of the emission region with respect to the central engine of GRB,
$E_{\rm ch}=1$~keV,
$f({E_{\rm ch}})={N_0}{K_2}\Delta T$,
$\Delta T$ is the duration of a pulse (see the 2nd column in Table~{\MyTabA}),
$\Gamma$ is the Lorentz factor of the emission region,
$F(\beta)\approx 0.597(-\beta)^{-2.30}$ for $-2.90 \leq\beta \leq -1.0 $,
and $c$ is the velocity of light.
The relation $R\simeq\Gamma^2 c \delta t/(1+z)$ is valid for the internal shock model,
where $\delta t$ is the full width at half maximum of the pulse in our analysis.
Setting $\tau_{\gamma\gamma } (E_{c})= 1$, the Lorentz factor $\Gamma $ is given by
\begin{equation}
\Gamma =[\sigma_T (\frac{d_L}{c \delta t})^2 E_{\rm ch} f(E_{\rm ch}) F(\beta)
(1+z)^{-2(\beta +1)} (\frac{E_{\rm c} E_{\rm ch}}{m_e^2
c^4})^{-\beta-1}]^{1/(2-2\beta)}.
\end{equation}

One should note that GRB spectrum usually appears to be a steep slope
and thus we only expect high-energy photons might be totally attenuated
by low-energy ones, rather than the other way around (\citealp{Li_Z-2010-ApJ.709.525L}).
Then, one can have ${E_{\rm c}}(1 + z) \gtrsim ( \Gamma {m_e}c^2 )^2/[ E_{\rm c}(1 + z)]$, or,
\begin{equation}
\Gamma \gtrsim \frac{E_{\rm c}}{m_e c^2}(1+z).
\end{equation}
In this paper, the Lorentz factor of a jet corresponding to a pulse with Band+cutoff radiation spectrum
is estimated with
\begin{equation}\label{Eq_Gamma}
\Gamma =\min \left\{ [\sigma_T (\frac{d_L}{c \delta t})^2 E_{\rm ch} f(E_{\rm ch}) F(\beta)
(1+z)^{-2(\beta +1)} (\frac{E_{\rm c} E_{\rm ch}}{m_e^2
c^4})^{-\beta-1}]^{1/(2-2\beta)}, \frac{E_{\rm c}}{m_e c^2}(1+z)\right \}.
\end{equation}

We note that the radiation spectrum of some pulses can be well described
with Band function rather than Band+cutoff.
That is to say that the high-energy spectrum shows no cutoff.
It implies that the optical depth of the maximum observed photon $E_{\max}$ by \emph{Fermi} satellite is
$\tau_{\gamma \gamma}(E_{\max})<1$.
This allows the minimum Lorentz factor to be estimated by
\begin{equation}\label{Eq_Gamma_low}
\Gamma>\Gamma_{\uparrow}=\min \left\{ [\sigma_T (\frac{d_L}{c \delta t})^2 E_{\rm ch} f(E_{\rm ch}) F(\beta)
(1+z)^{-2(\beta +1)} (\frac{E_{\rm max} E_{\rm ch}}{m_e^2
c^4})^{-\beta-1}]^{1/(2-2\beta)}, \frac{E_{\rm max}}{m_e c^2}(1+z)\right \}.
\end{equation}
Equation~(\ref{Eq_Gamma_low}) is used to estimate the lower limit of Lorentz factor for pulse with Band radiation spectrum.

\emph{Lorentz factor for pulses and Associated Correlations}.$\;\;$
Based on our spectral fitting results and Equation~(\ref{Eq_Gamma}) or (\ref{Eq_Gamma_low}),
we estimate the Lorentz factor or its lower limit for the jet corresponding to our pulses,
which are reported in Table~{\MyTabB}.
The Lorentz factor $\Gamma$ (blue symbols) or its lower limit $\Gamma_{\uparrow}$ (green symbols)
are also plotted in Figure~{\MyFigA}. It can be found that the Lorentz factor of pulses with high-energy spectral cutoff in our sample ranges from 60 to 682.

It has been suggested that the Lorentz factor correlates with other quantities of the pulses
such as the value of $L_{\rm \gamma,iso}$ or $E_{\rm p,z}$ (e.g., \citealp{Lu_J-2012-Zou_YC-ApJ.751.49L,Liang_EW-2015-Lin_TT-ApJ.813.116L}).
\cite{Lu_J-2012-Zou_YC-ApJ.751.49L} makes a detail analysis of 38 GRBs with
the initial bulk Lorentz factor of the jet producing the external-shock afterglow
and found a tight relation between $\Gamma_{\rm 0}$ and $L_{\rm \gamma,iso}$, i.e., $\Gamma_{\rm 0} \simeq 249 L_{\rm \gamma,iso,52}^{0.30}$.
In the upper-left panel of Figure~{\MyFigC}, we show the relation of $\Gamma$-$L_{\rm \gamma,iso}$ for our pulses,
where the blue and green symbols represent the quantities of the pulse with the exact or lower limit of Lorentz factor,
the black line is the relation of $\Gamma_{\rm 0} \simeq 249 L_{\rm \gamma,iso,52}^{0.30}$,
and the black and red symbols are the same as those in figure~2 of \cite{Lu_J-2012-Zou_YC-ApJ.751.49L}.
It is found that the relation of our $\Gamma$-$L_{\rm \gamma,iso}$ is
consistent with $\Gamma_{\rm 0} \simeq 249 L_{\rm \gamma,iso,52}^{0.30}$ but with slightly high index.

It is found that the Lorentz factor $\Gamma$ of the outflow
not only depends on the luminosity $L_{\rm \gamma,iso}$, but also depends on $E_{\rm p,z}$.
\cite{Liang_EW-2015-Lin_TT-ApJ.813.116L} compiled a sample of 34 long GRBs
with known $L_{\rm \gamma,iso}$, $E_{\rm p,z}$, and $\Gamma_{0,\rm es}$
and studied the relation among $L_{\rm \gamma,iso}$, $E_{\rm p,z}$, and $\Gamma_{0,\rm es}$.
The relations of $\log L_{\rm iso, 52}=-(6.38\pm0.35)+(1.34\pm0.14)\times \log (E_{\rm p,z}/{\rm keV})+(1.32\pm0.19)\times \log\Gamma_{0,\rm es}$
and $\log (E_{\rm p,z}/{\rm keV})=(3.71\pm0.38)+(0.55\pm0.06)\times \log L_{\rm iso, 52}-(0.50\pm0.17)\times \log \Gamma_{0,\rm es}$
are obtained.
In the upper-right and bottom-left panels of Figure~{\MyFigC},
we show the relations of $L^{\rm r}_{\rm iso}$-$L_{\rm iso}$ and $E_{\rm p,z}^{\rm r}$-$E_{\rm p,z}$ for our pulses,
where the values of $L_{{\rm{iso}},{\rm{52}}}^{\rm{r}}$ ($E^{\rm r}_{\rm p,z}$) for our pulses is estimated with the formula $L_{{\rm{iso}},{\rm{52}}}^{\rm{r}} = {10^{ - 6.38}}{({E_{{\rm{p}},{\rm{z}}}}/{\rm{keV}})^{1.34}}{\Gamma ^{1.32}}$ ($E^{\rm r}_{\rm p,z}=10^{3.71}L^{0.55}_{\rm iso, 52}\Gamma^{-0.50}$~keV) and the $L_{\rm iso, 52}$($E_{\rm p,z}$) is obtained through observational data.
The black symbols, solid lines, and dashed lines are the same as those in the upper-left and upper-right panels of figure~2 in \cite{Liang_EW-2015-Lin_TT-ApJ.813.116L},
the blue and green symbols are the same as those in Figure~{\MyFigA}.
These two panels reveal that our pulses are highly consistent with correlations reported in \cite{Liang_EW-2015-Lin_TT-ApJ.813.116L}.

\emph{Lorentz factor for External-shock Afterglow}.$\;\;$
Central engines of GRBs may be intermittent and launch several episodes of jet shells,
which is responsible for different pulses in the prompt emission.
In this scenario, the external-shock is formed due to the propagation
of the preceding (merged) jet shell
into the circum-burst medium.
The jet shells launched from the central engine would collide with each other
and a single merged jet shell may appear soon afterwards.
If the onset bump of the afterglow appears after the prompt emission,
the single merged shell is responsible for the onset bump.
In this scenario, the initial Lorentz factor ${\Gamma _0}$ of the jet producing the external-shock afterglow can be estimated as
${\Gamma _0}=\Gamma_{0,\rm ps}$ with
\begin{equation}\label{Eq:Gamma_Esta}
\Gamma_{0,\rm ps}= \frac{{\sum\limits_{n = 1}^N {\frac{{1 - {\eta _{{\rm{rad}},n}}}}{{{\eta _{{\rm{rad}},n}}}}{E_{\gamma ,{\rm{iso}},n}}} }}{{\sum\limits_{n = 1}^N {\frac{{1 - {\eta _{{\rm{rad}},n}}}}{{{\eta _{{\rm{rad}},n}}}}\frac{{{E_{\gamma ,{\rm{iso}},n}}}}{{{\Gamma _n}}}} }} = \frac{{\sum\limits_{n = 1}^N {{E_{\gamma ,{\rm{iso}},n}}} }}{{\sum\limits_{n = 1}^N {\frac{{{E_{\gamma ,{\rm{iso}},n}}}}{{{\Gamma _n}}}} }},
\end{equation}
where ${{E_{\gamma ,{\rm{iso}},n}}}$, ${{\Gamma _n}}$, and ${{\eta _{{\rm{rad}},n}}}$ are the isotropic energy, Lorentz factor, and the radiation efficiency of the jet corresponding to $n$th pulse in a burst.
For our bursts, the value of $\Gamma_{0,\rm ps}$ is estimated and compared with the initial Lorentz factor $\Gamma _{0,\rm es}$ estimated based on the external-shock afterglow,
which are shown in the bottom-right panel of Figure~{\MyFigC}.
The values of $\Gamma_{0,\rm ps}$ and $\Gamma _{0,\rm es}$ for the burst in our sample are also reported in Table~{\MyTabC}.
The detailed discussion about the values of $\Gamma_{0,\rm ps}$ and $\Gamma _{0,\rm es}$ will be made in Section~\ref{Sec:bursts discussion}.

\subsection{Lorentz Factor Evolution for an individual Burst in our sample}\label{Sec:bursts discussion}
According to the results showed in Figure~{\MyFigA},
one can find that the Lorentz factor of pulses with high-energy spectral cutoff in our sample ranges from 60 to 682.
In addition, the Lorentz factor of jets in an individual burst generally fluctuates
within a certain range and without a general trend, except GRBs~130821A, 160509A, and 160625B.
In GRBs~130821A, 160509A, and 160625B, the Lorentz factor of jets in an individual burst seems to increase with time.

\begin{itemize}
\item
The prompt emission of GRB~090323, which is the second most energetic LAT-detected burst after GRB~080916C,
consists with multiple distinct pulses over $\sim150$~s and a main emission episode in the period of $[0, 70]$~s.
With the cutoff energy in the radiation spectrum,
the Lorentz factor of jet corresponding to different pulses
can be found in Figure~{\MyFigA}.
Based on the assumption that the peak flux time in the LAT light-curve
($\sim$40~s) represents the fireball deceleration
time, \cite{Ackermann_M-2013-Ajello_M-ApJS.209.11A} obtained $\Gamma_{0,\rm es}\sim 350-870$ in an ISM circum-burst environment and $\Gamma_{0,\rm es}\sim 350-590$ in a wind circum-burst environment.
From Figure~{\MyFigA}, only one pulse appears at $t_{\rm obs}<40$~s
and the corresponding Lorentz factor is estimated to be 186,
which is very different from $\Gamma_{0,\rm es}$ estimated in \cite{Ackermann_M-2013-Ajello_M-ApJS.209.11A}.
In the scenario that the pulses in this burst formed in the jets launched in different times,
the bulk Lorentz factor of the jet producing the external-shock afterglow at $t_{\rm obs}\lesssim 70$~s is estimated
to be $\Gamma_{0,\rm ps}\gtrsim 183.45$ based on Equation~(\ref{Eq:Gamma_Esta}).
Unfortunately, the Lorentz factor of the fireball in the afterglow phase at $t_{\rm obs}\gtrsim 70$~s
is no estimated.
\item
GRB~090926A is a long bright GRB and clearly shows a short spike at $\sim 10$~s in all detectors of \textsl{Fermi} satellite (\citealp{Ackermann_M-2011-Ajello_M-ApJ.729.114A}).
However, the high-energy spectral cutoff is only found in two pulses of this burst.
Then, the evolution of Lorentz factors could not be constrained.
Based on the assumption that the peak flux time in the LAT light-curve ($\sim 10$~s) represents the fireball deceleration
time, \cite{Ackermann_M-2013-Ajello_M-ApJS.209.11A} estimate the initial Lorentz factor of the jet producing the external-shock afterglow.
The value of $\Gamma_{0,\rm es}\sim 520-700$ and $\sim 400-450$ are obtained
for an ISM and a wind circum-burst environment, respectively.
For this burst, the initial Lorentz factor of the jet in the afterglow phase
is also estimated to be $\Gamma_{0,\rm ps}\gtrsim 304$ based on Equation~(\ref{Eq:Gamma_Esta}).
This value is consistent with the value of $\Gamma_{0,\rm es}$ in a wind environment as shown
in the bottom-right panel of Figure~{\MyFigC},
which may imply a wind environment for this burst.
Similarly, the high-energy spectral cutoffs are also found only in two pulses in GRB~120226A and GRB~170405A.
Then, the evolution of Lorentz factor for pulses could not be constrained in these two bursts.
\item
GRB~100724B has multiple peaks of varying intensity as shown in Figure~{\MyFigA}.
Characteristic features of GRB~100724B are the simultaneous emissions at MeV and GeV (\citealp{DelMonte_E-2011-Barbiellini_G-A&A.535A.120D}).
This GRB displays a pronounced Lorentz factor evolution pattern,
with its evolution beginning slowly and then varying around $100$ within a small fluctuation.
For GRB~100724B, the initial Lorentz factor of the jet in the external-shock phase is estimated as $\Gamma_{0,\rm ps}\gtrsim 105$ based on Equation~(\ref{Eq:Gamma_Esta}).
Unfortunately, there is no afterglows used to estimated the initial Lorentz factor of jet producing the external-shock.
\item
GRB~130821A can be decomposed into the main emission episode,
which is presented as an obvious multi-peak lasting around 40~s
and contributes around $82.43\%$ of the $\gamma$-rays energy in this burst,
and the followed two sporadic short peaks at $\sim 55$~s and $90$~s, respectively.
Figure~{\MyFigA} reveals that the Lorentz factor of GRB~130821A in the main emission episode seems to increase with time.
With the peak flux time of the LAT light-curve,
the initial bulk Lorentz factor of jet is estimated to be $\Gamma_{0,\rm es}\sim440$ (\citealp{Liang_YF-2014-Zhou_B-ApJ.781.74L}),
which is very different from the value estimated based on Equation~(\ref{Eq:Gamma_Esta}),
 i.e., $\Gamma_{0,\rm ps}\gtrsim 99$.
The deviation between $\Gamma_{0,\rm ps}$ and $\Gamma_{0,\rm es}$ can be solved
if the bulk Lorentz factor of jet producing the main emission episode in GRB~130821A
increases with time.
Together with the Lorentz factor evolution shown in Figure~{\MyFigA},
we would like to believe that the Lorentz factor of the jet or jets in GRB~130821A
should increase with time.
\item
GRB~160509A consists of multi-pulses with Lorentz factor of jets fluctuating around 180.
With \emph{Swift} X-ray and ground-based radio, near-infrared, and optical data up to 20 days,
\cite{Laskar_T-2016-Alexander_KD-ApJ.833.88L} argued that
the afterglow emission comprises distinct external-reverse shock and external-forward shock contributions.
Based on the joint analysis of these two emission components,
they finally derived the initial Lorentz factor $\Gamma_{0,\rm es}\sim330$ based on the deceleration time $\sim T_{90}$.
Taking Fermi-LAT observations into account, however,
\cite{Fraija_N-2020-Laskar_T-ApJ.905.112F}
obtained an initial Lorentz factors of $\Gamma_{0,\rm es}\sim600$
(corresponding to the deceleration time $\sim88$~s for a fireball with $E_{\rm k,iso}=6.98\times 10^{53}$~erg decelerated in an uniform-density circum-burst medium $n=4.56\times 10^{-4}\;\rm cm^{-3}$)
after modeling the multi-wavelength observations together with a external-forward and external-reverse shock.
For this burst, the initial Lorentz factor of the jet is estimated as $\Gamma_{0,\rm ps}\sim 186$ for pulses appearing at $t_{\rm obs}\lesssim 20$~s based on Equation~(\ref{Eq:Gamma_Esta}).
The deviation between $\Gamma_{0,\rm ps}$ and $\Gamma_{0,\rm es}$
can be solved if the bulk Lorentz factor of jet in GRB~160509A increases with time.
In Figure~{\MyFigA}, the Lorentz factor of jets corresponding to different pulses in GRB~160509A indeed seems to be increase with time.
\item
GRB~160625B is one of the brightest bursts in recent years.
One distinct feature of this burst is the multi-bursting behavior,
which is separated by two quiescent times.
\cite{Alexander_KD-2017-Laskar_T-ApJ.848.69A} argued that the radio emission of this burst
is dominated by the external-reverse shock and external-forward shock components formed during the jet propagating in an ISM environment, and adopt three external-reverse shock models to decipher the radio data.
The corresponding bulk Lorentz factor of the jet producing the external-reverse shock are collected in Table~{\MyTabC}.
For this burst, the initial Lorentz factor of the jet producing the external-shock afterglow
is estimated to be $\Gamma_{0,\rm ps}\gtrsim 240$ based on Equation~(\ref{Eq:Gamma_Esta}),
which is consistent with $\Gamma_{0,\rm es}\sim290$ estimated based on the model~1 in \cite{Alexander_KD-2017-Laskar_T-ApJ.848.69A}.
However, \cite{Fraija_N-2017-Veres_P-ApJ.848.15F} argued that the early afterglow are consistent
with the external-shock in a wind-like circum-burst environment
and the late afterglow are consistent with the external-shock in an ISM circum-burst environment.
They obtained the initial Lorentz factor $\Gamma_{0,\rm es}\sim500$,
which is corresponding to the deceleration time $\sim225$~s
based on equation~(46) of \cite{Fraija_N-2015-ApJ.804.105F}.
This may reveal that the Lorentz factor of the jet or jets in GRB~160625B may increase with time in the episode of $[185, 225]$~s if this burst is in the wind-to-ISM circum-burst environment.
\item
GRB~180720B is a long and powerful burst with observation of sub-TeV gamma-rays in its afterglow.
The simultaneous multi-wavelength observations of this burst were presented
over multiple periods of time beginning just after the trigger time and extending to more than 30 days (\citealp{Fraija_N-2019-Dichiara_S-ApJ.885.29F}).
The prompt emission of this burst last around 50~s,
with main emission episode in the period of $[0, 20]$~s.
Figure~{\MyFigA} reveals that the Lorentz factor of GRB~180720B in the main emission episode varies with time
and without an increase or a decrease trend.
Based on the assumption that the peak flux time in the LAT light-curve represents the fireball deceleration time,
\cite{Ronchi_M-2020-Fumagalli_F-AA.636A.55R} estimated the initial Lorentz factor of the jet producing the external-shock afterglow with a wind-like or ISM circum-burst environment.
The results are collected in Table~{\MyTabC}.
Based on Equation~(\ref{Eq:Gamma_Esta}),
the initial Lorentz factor $\Gamma_{0,\rm ps}$ of the jet is estimated to be $\gtrsim76$,
which is consistent with that estimated in the wind circum-burst environment, i.e., $\Gamma _{0,\rm es}\sim80$ (\citealp{Ronchi_M-2020-Fumagalli_F-AA.636A.55R}).
This result may reveal that the jet launched in different episode is responsible for the different pulse in the main emission episode of this burst.
\end{itemize}

\section{Conclusions}\label{Sec:Conclusions}
This paper is dedicated to study the Lorentz factor evolution for jets in an individual GRB during the process of the burst.
Then, we search for GRBs with spectral cutoff featuring in their high-energy regime during its prompt emission phase.
We showed nine GRBs (GRBs~090323, 090926A, 100724B, 120226A, 130821A, 160509A, 160625B, 170405A, and 180720B)
that satisfy our sample selection principles.
A total of 70 different pulses is obtained, 34 of which are characterized by high-energy spectral cutoff.
In the scenario that the two-photon pair production is responsible for high-energy spectral cutoff,
we estimated the Lorentz factor of jet corresponding to the studied pulse.
It is found that the Lorentz factor of jet corresponding to
the pulse with high-energy spectral cutoff in our sample ranges from 60 to 682.
In addition, the Lorentz factor of jets in an individual burst generally fluctuates
within a certain range and without a general trend,
except GRB~130821A, GRB~160509A and GRB~160625B. In GRB~130821A, GRB~160509A and GRB~160625B,
the Lorentz factor of jets in an individual burst seems to increase with time.
In GRBs~090926A, 120226A and 170405A, the high-energy spectral cutoff is only found in two pulses of these bursts
and thus the Lorentz factor evolution in these bursts could not be constrained.
The different behavior of the Lorentz factor evolution found in these GRBs may
suggest the difference in jet composition in these bursts.
Meanwhile, we also examine the relations among $L_{\rm iso }$, $E_{\rm p,z}$, and $\Gamma$ for pulses in our sample.
It is found that the $L_{\rm iso }-\Gamma$ relation is consistent with that found in \cite{Lu_J-2012-Zou_YC-ApJ.751.49L}
and the $L_{\rm iso }-E_{\rm p,z}-\Gamma$ relations are consistent with these reported in
\cite{Liang_EW-2015-Lin_TT-ApJ.813.116L}.

\acknowledgments
We acknowledge the use of the Fermi archive's public data.
This work is supported by the National
Natural Science Foundation of China (grant Nos. 11773007,
11673006, U1938116, U1938201, U1731239, and U1938106),
the Guangxi Science Foundation (grant Nos. 2018GXNSFFA281010,
2017AD22006, 2018GXNSFGA281007, and 2018GXNSFDA281033).

\clearpage
\begin{figure}
\begin{tabular}{ccc}
\includegraphics[angle=0,scale=0.23]{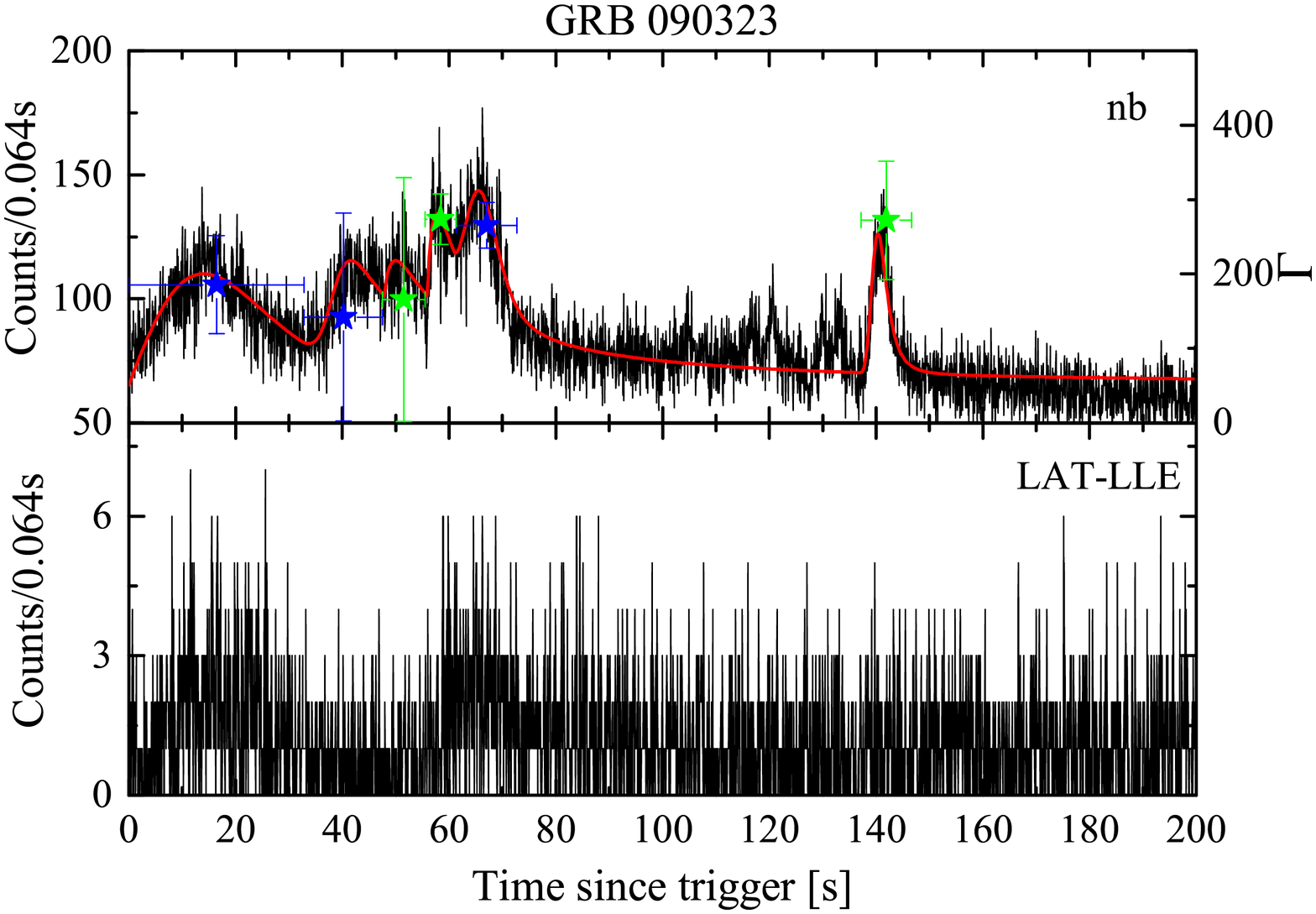} &
\includegraphics[angle=0,scale=0.23]{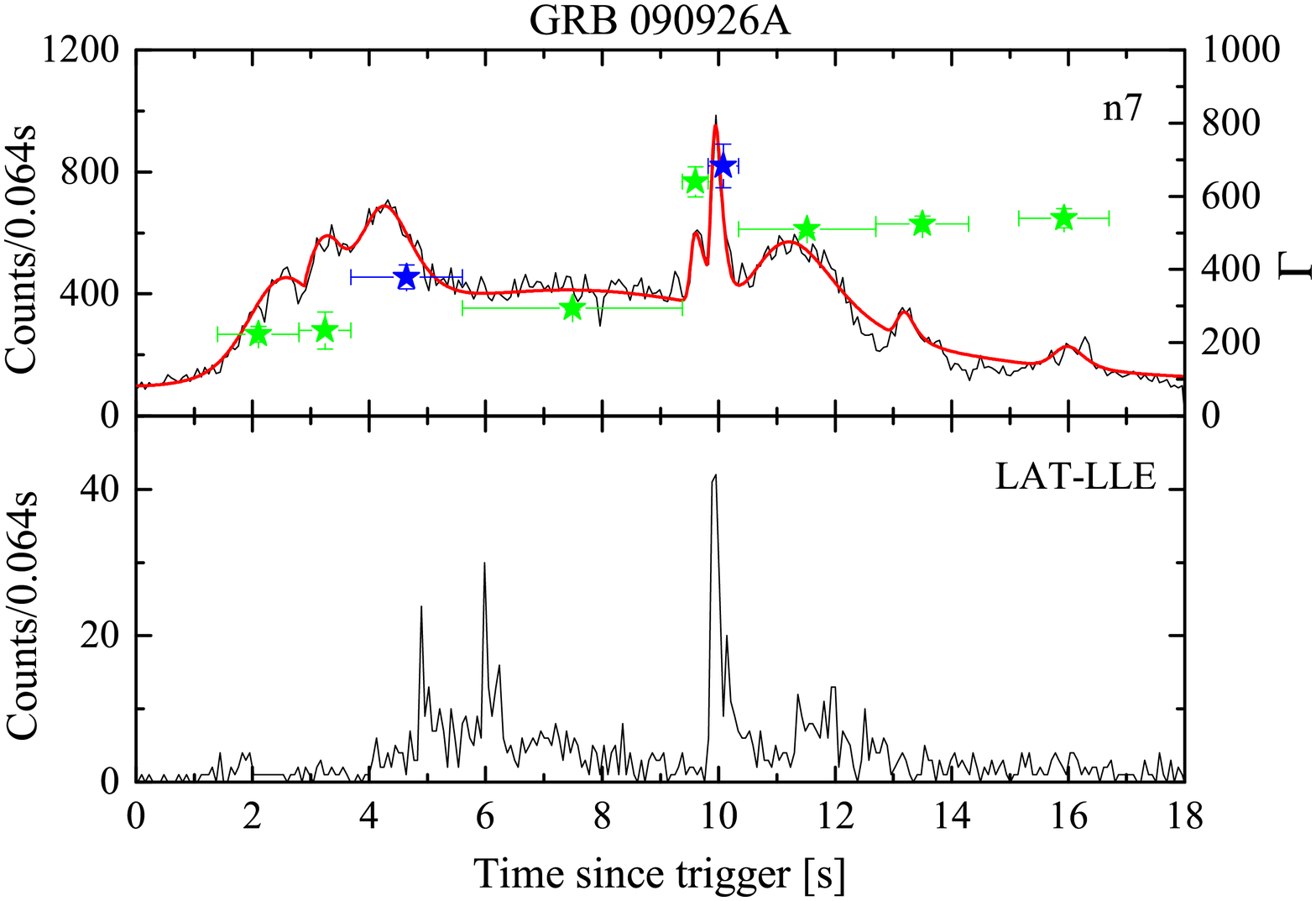} \\
\includegraphics[angle=0,scale=0.23]{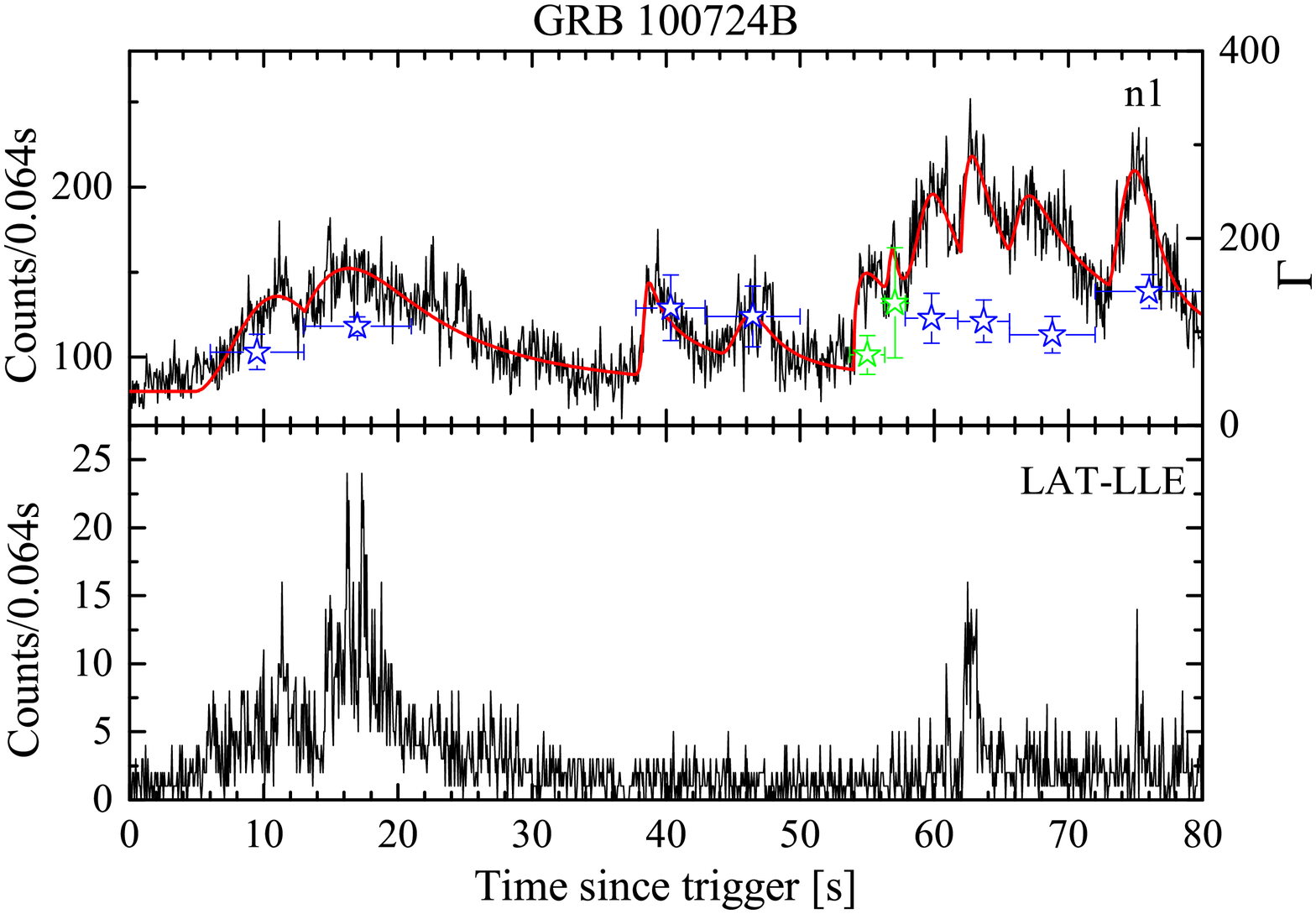} &
\includegraphics[angle=0,scale=0.23]{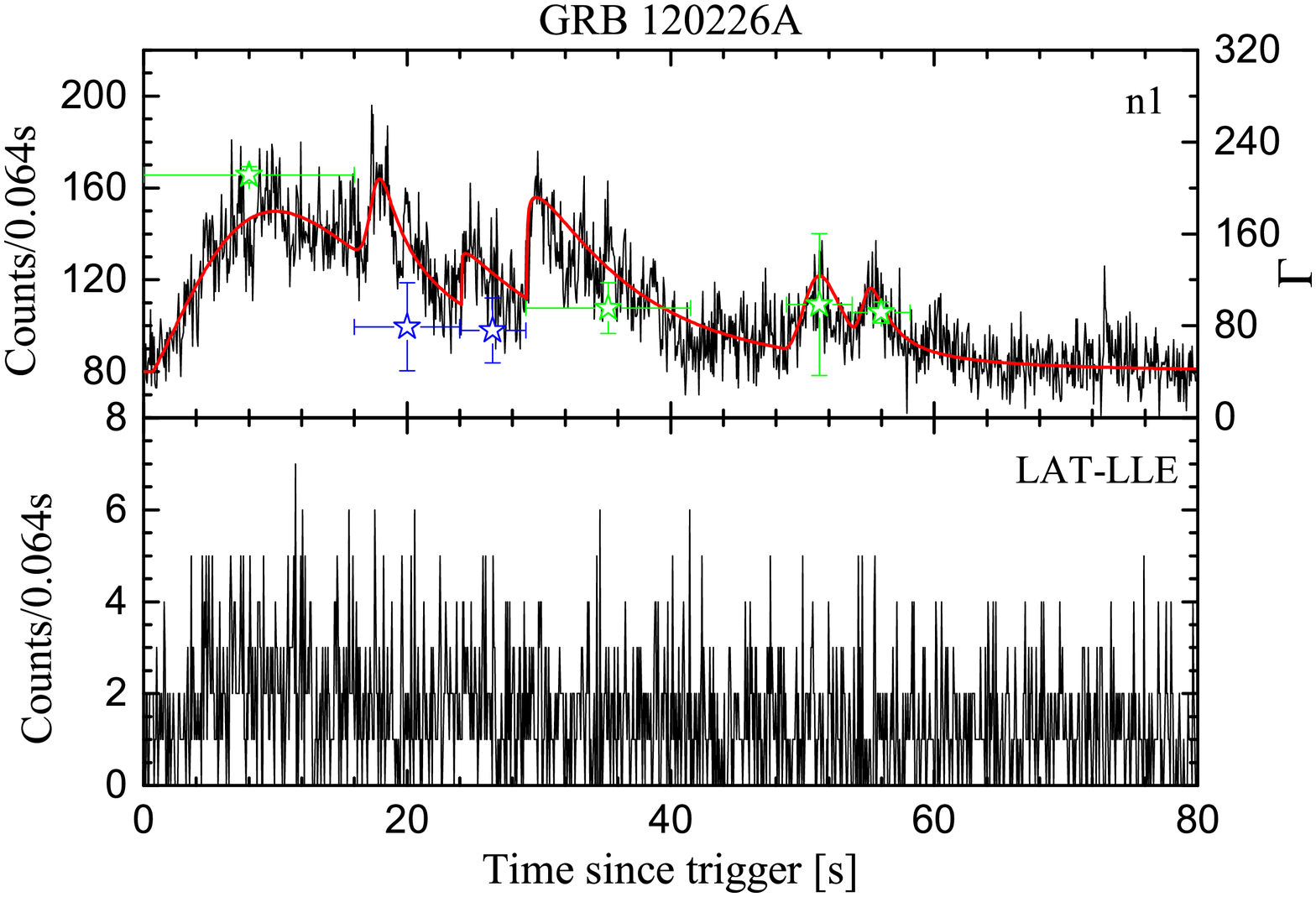} \\
\includegraphics[angle=0,scale=0.23]{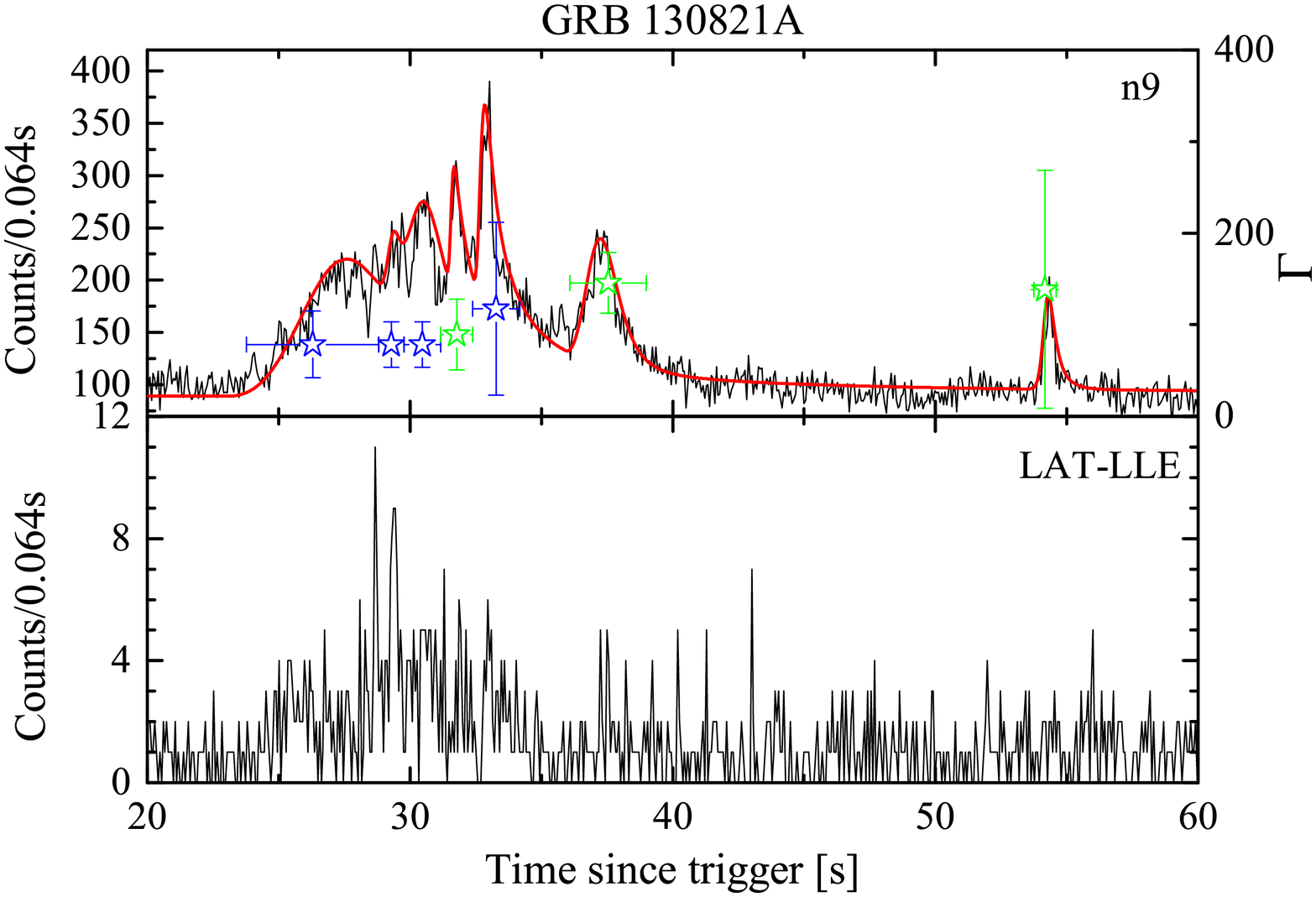} &
\includegraphics[angle=0,scale=0.23]{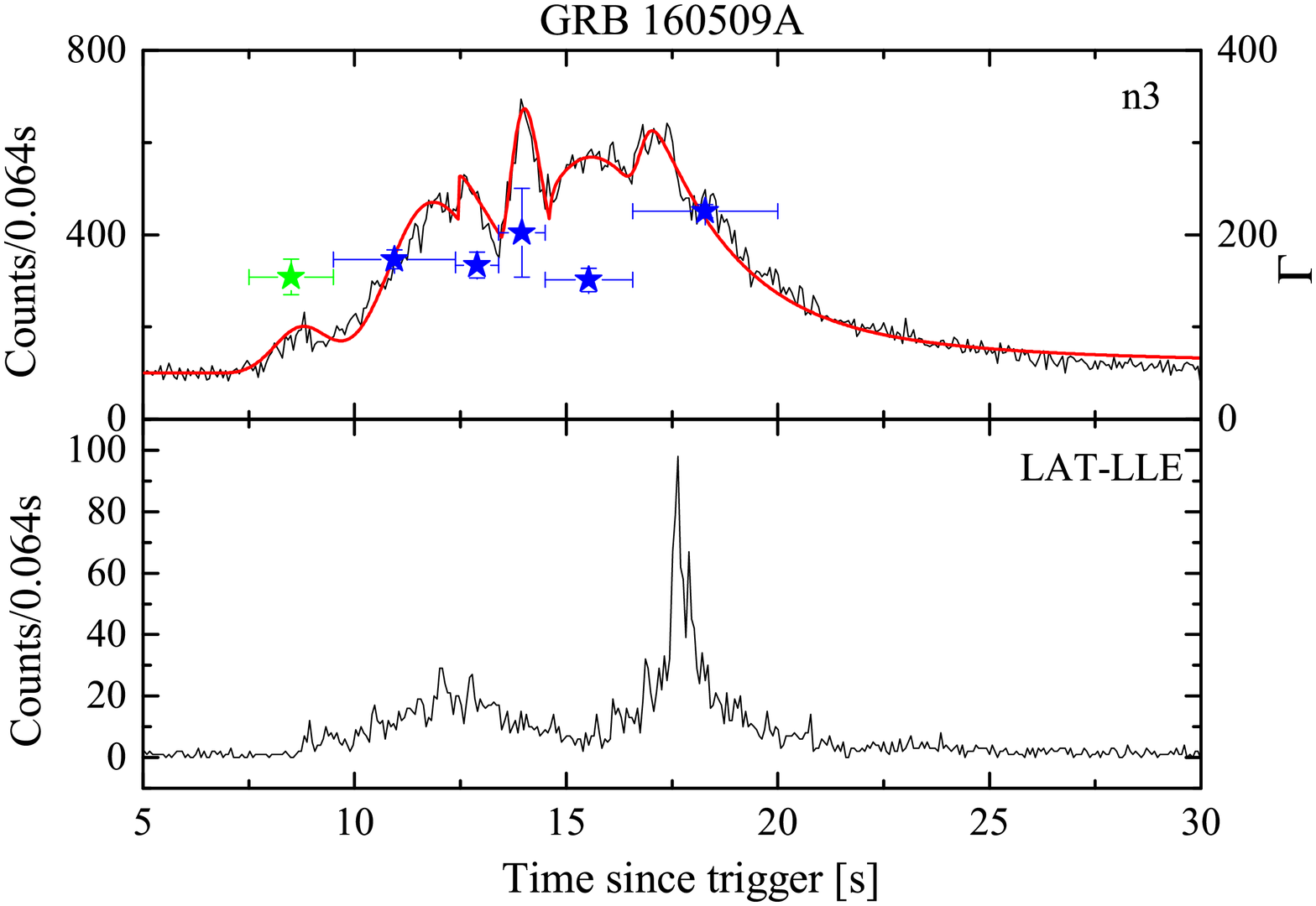} \\
\includegraphics[angle=0,scale=0.23]{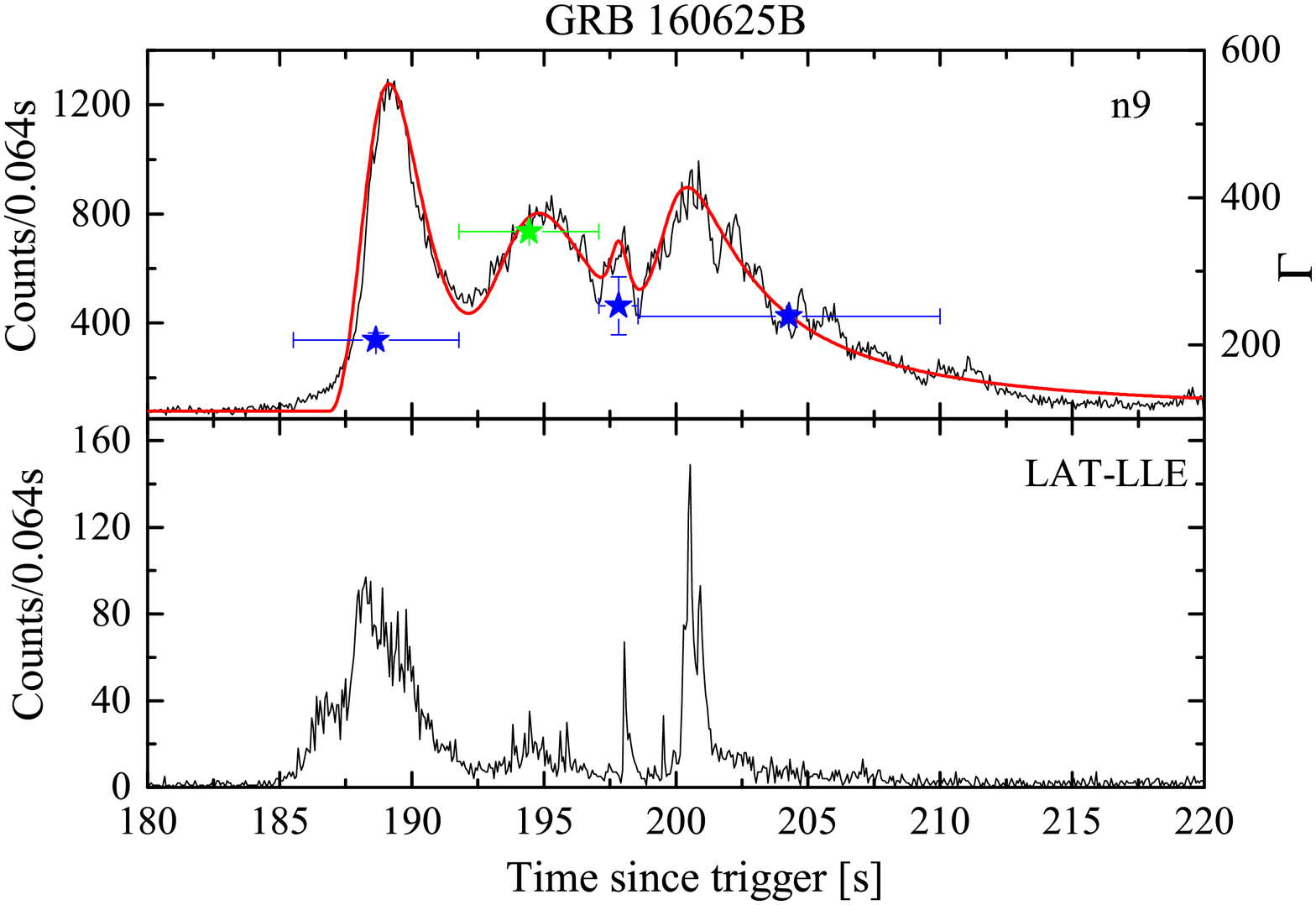} &
\includegraphics[angle=0,scale=0.23]{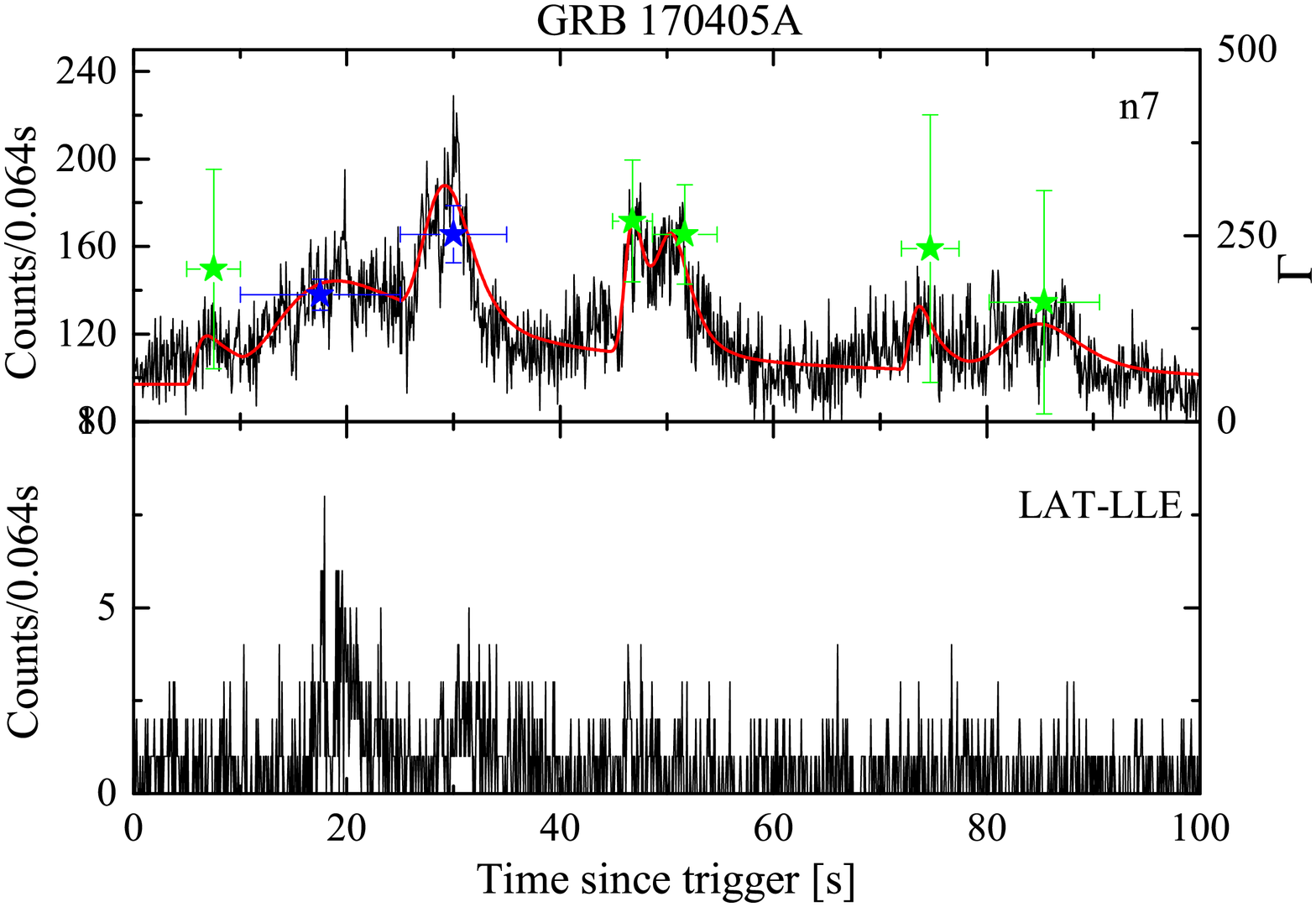} \\
\includegraphics[angle=0,scale=0.23]{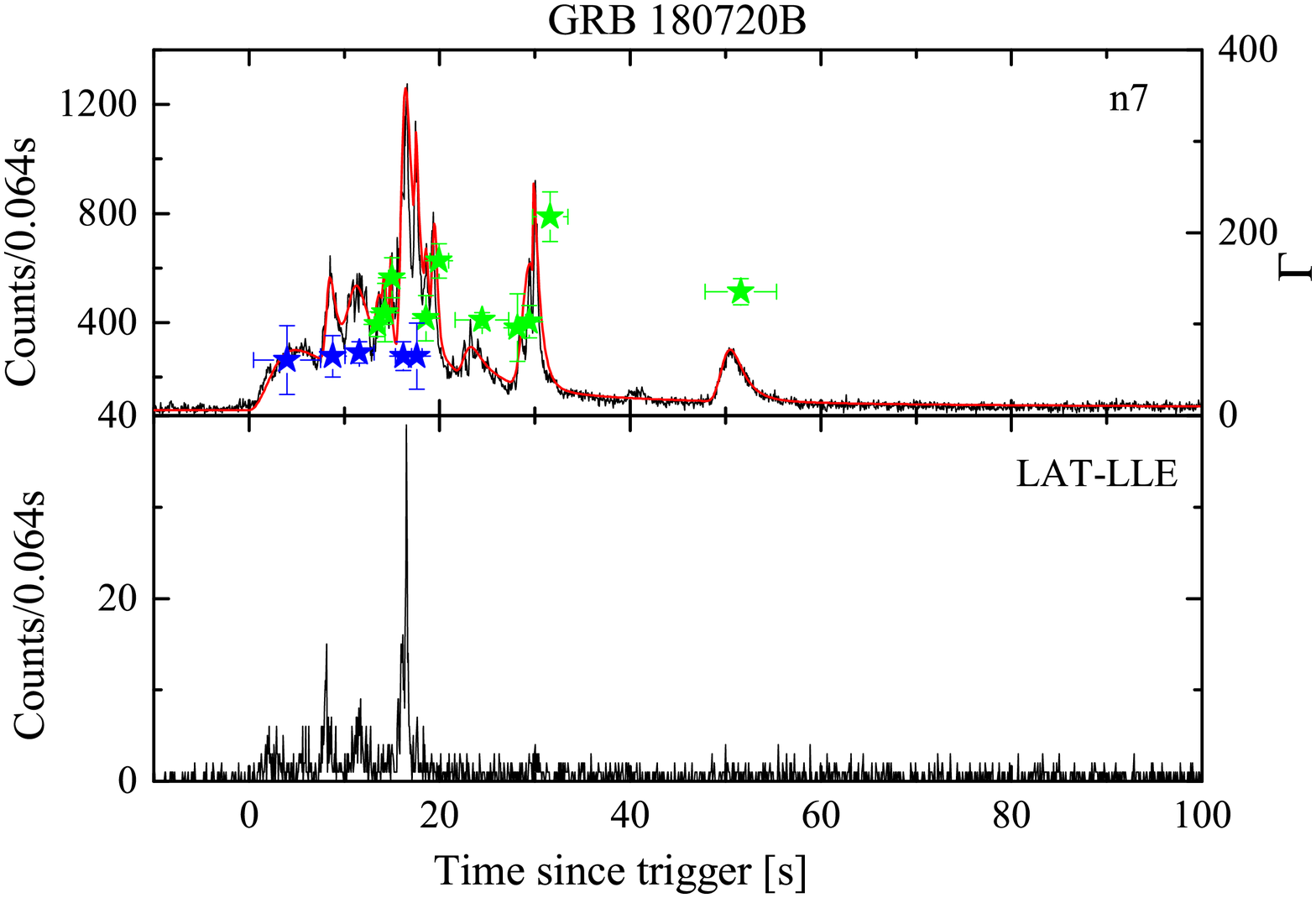} &
\end{tabular}
\caption{The light-curves (black lines) and evolution of Lorentz factor $\Gamma$ (blue symbols)
or its low limit $\Gamma_{\uparrow}$ (green symbols) of our sample, where the red lines represent the light-curve fitting with Equation~(\ref{Eq:LC_Fitting}). Solid and hollow ``$\bigstar$'' are those with and without redshift detection, respectively;
for bursts without redshift detection, we use redshift $z=1.0$.}\label{MyFigA}
\end{figure}
\clearpage
\begin{figure}
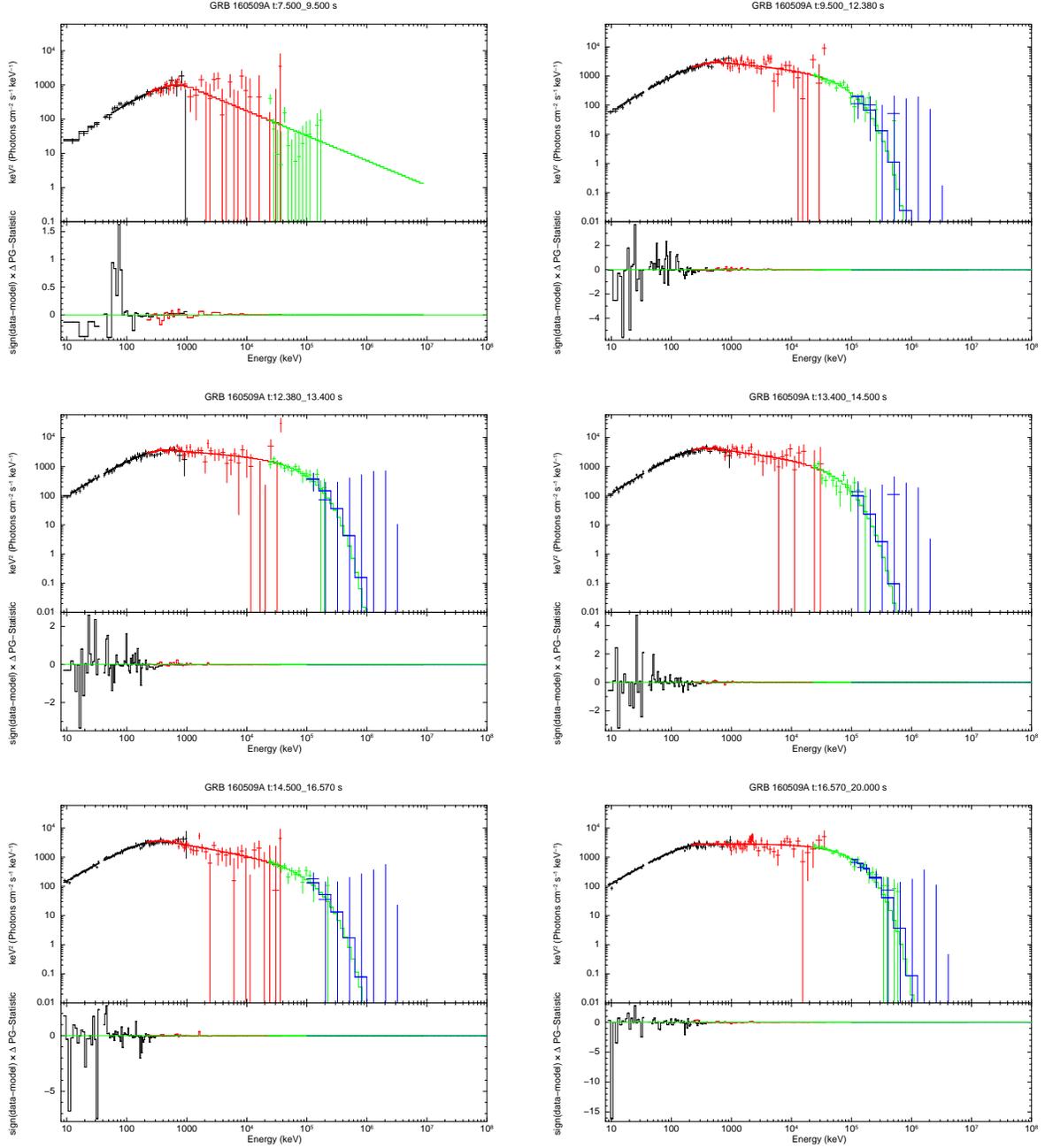

\begin{tabular}{cc}
\includegraphics[angle=270,scale=0.30]{Fig_160509A_01.eps} &
\includegraphics[angle=270,scale=0.30]{Fig_160509A_02.eps} \\
\includegraphics[angle=270,scale=0.30]{Fig_160509A_03.eps} &
\includegraphics[angle=270,scale=0.30]{Fig_160509A_04.eps} \\
\includegraphics[angle=270,scale=0.30]{Fig_160509A_05.eps} &
\includegraphics[angle=270,scale=0.30]{Fig_160509A_06.eps} \\
\end{tabular}
\caption{Spectrum fitting results for pulses in GRB~160509A.}\label{MyFigB}
\end{figure}

\clearpage
\begin{figure}
\begin{tabular}{cc}
\includegraphics[angle=0,scale=0.35, trim=60 10 40 0]{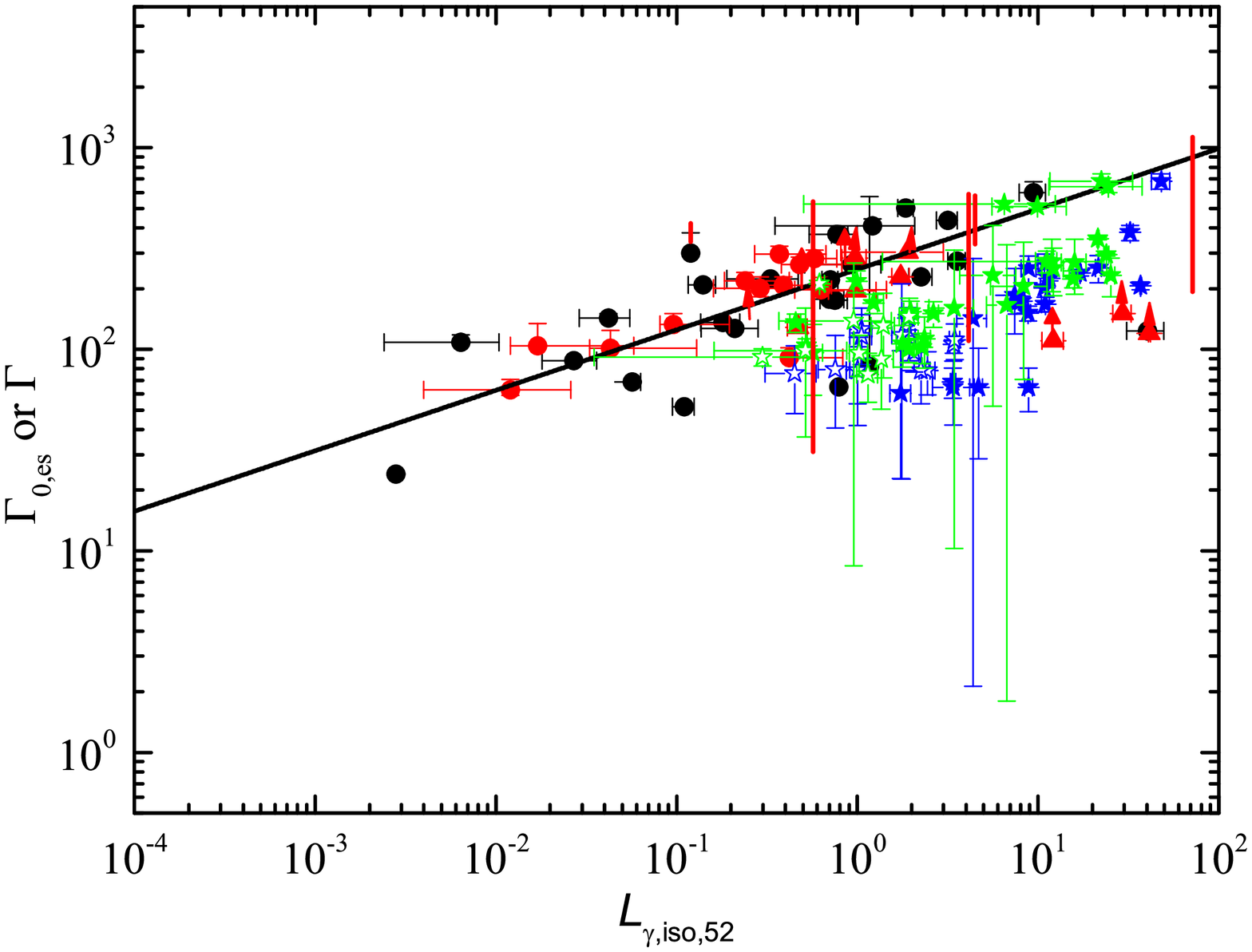} &
\includegraphics[angle=0,scale=0.33, trim=60 0 40 0]{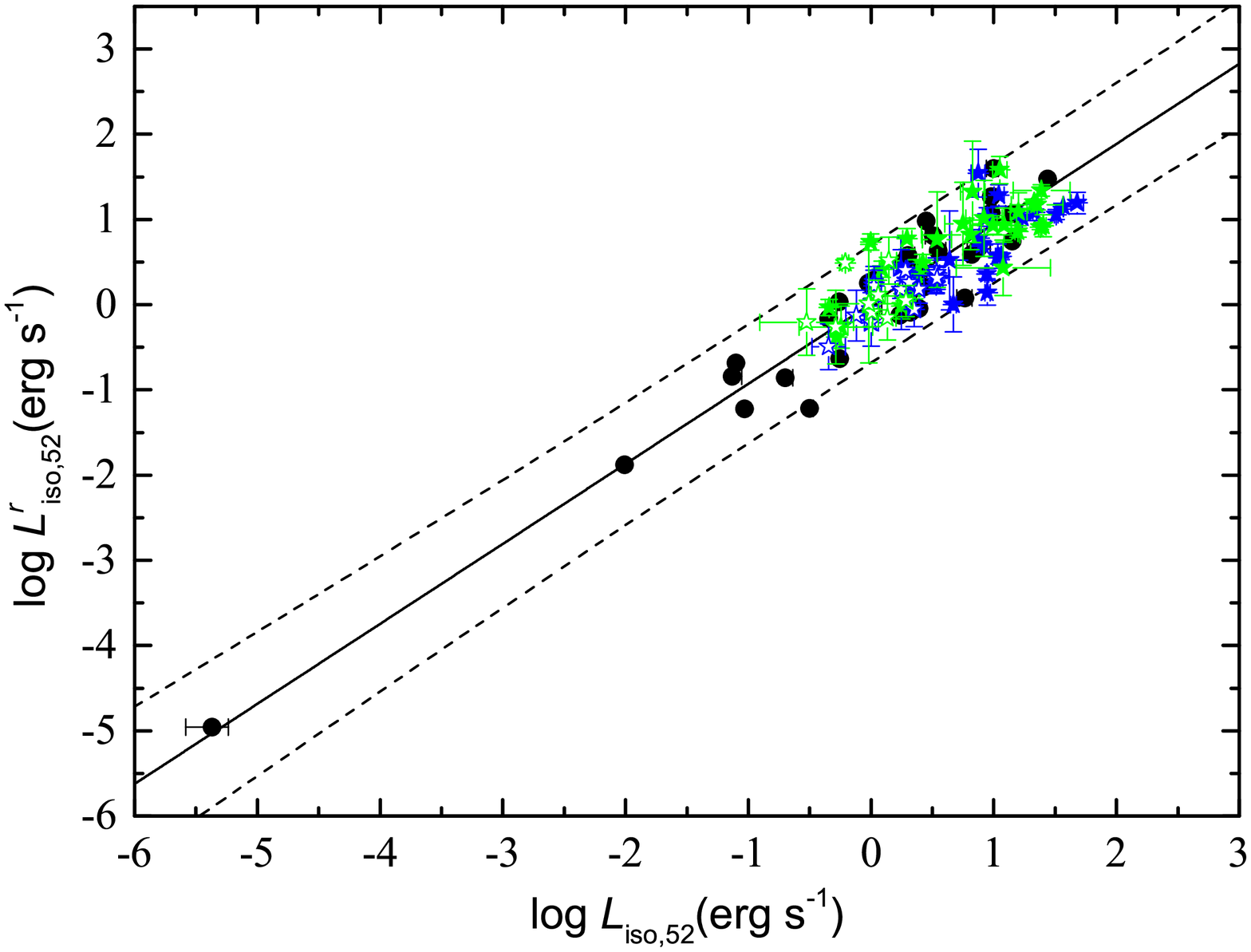} \\
\includegraphics[angle=0,scale=0.33, trim=60 0 40 0]{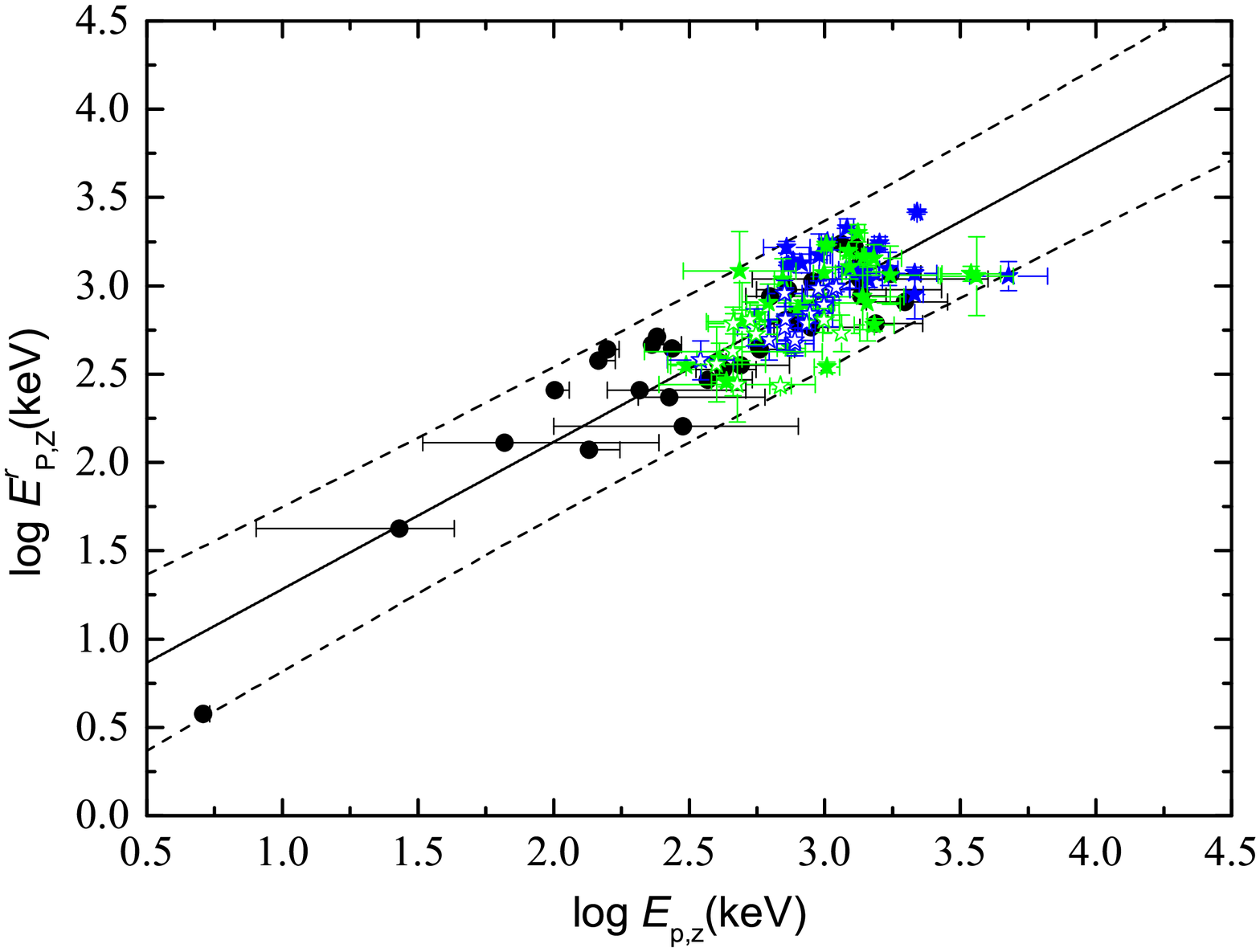} &
\includegraphics[angle=0,scale=0.33, trim=60 0 40 0]{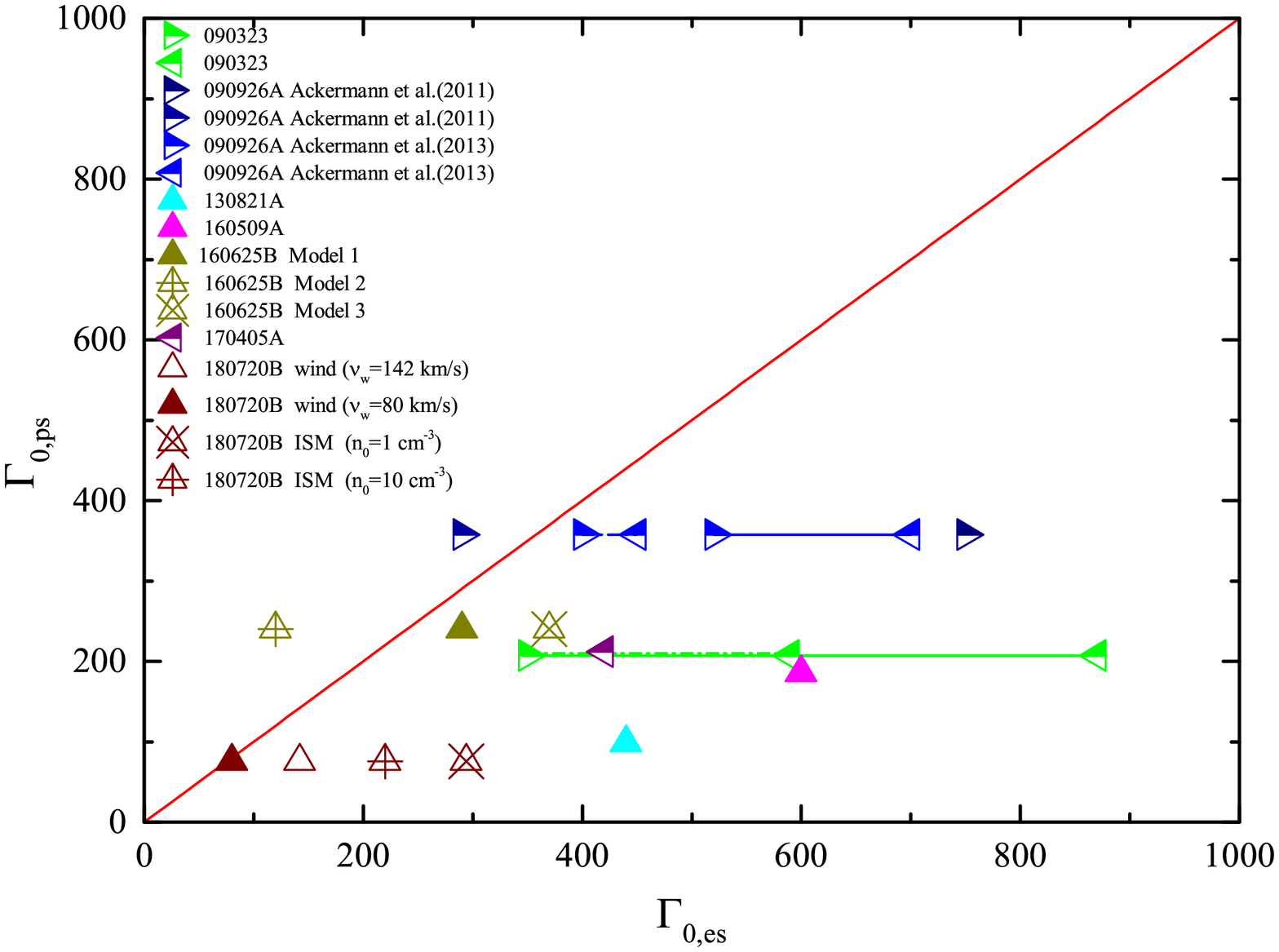} \\
\\
\end{tabular}
\caption{{\em Upper-left panel---} $\Gamma-L_{\rm \gamma,iso}$ relation for our pulses,
where the black solid line, black and red symbols are the same as those in figure~2 of \cite{Lu_J-2012-Zou_YC-ApJ.751.49L},
and the blue (green) symbols and solid (hollow) ``$\bigstar$'' are the same as those in Figure~{\MyFigA}.
 {\em Upper-right and bottom-left panels---} The relations of $L^{\rm r}_{\rm iso}$-$L_{\rm iso}$ and $E_{\rm p,z}^{\rm r}$-$E_{\rm p,z}$ for our pulses, where the black symbols, solid lines, and dashed lines are the same as those in the upper-left and upper-right panels of figure~2 in \cite{Liang_EW-2015-Lin_TT-ApJ.813.116L}, and the blue (green) symbols and solid (hollow) ``$\bigstar$'' are the same as those in Figure~{\MyFigA}.
{\em Bottom-right panel---} The value $\Gamma_{0,\rm ps}$ estimated based on Equation~(\ref{Eq:Gamma_Esta}) vs. $\Gamma _{0,\rm es}$ estimated based on the external-shock afterglow.
Dash dot and solid lines represent the wind circum-burst environment and ISM circum-burst environment, respectively.
In the panel, we only show the results of 0-150~s for GRB~090323 and 0-18~s for GRB~090926A.
}
\label{MyFigC}
\end{figure}

\begin{center}
\begin{deluxetable}{llllllllcl}
\tablewidth{0pt} \tabletypesize{\footnotesize}
\tabletypesize{\tiny}
\tablecaption{Light-curve fitting results for the pulse in our sample}
\tablehead{
\colhead{GRB}& \colhead{pulse}&
\colhead{$F_{\rm p}$\tablenotemark{\tiny a}}& \colhead{$t_0$\tablenotemark{\tiny b}}& \colhead{$t_{\rm p}$\tablenotemark{\tiny b}}& \colhead{$r$}& \colhead{$d$}& \colhead{$\delta t$\tablenotemark{\tiny b}}&\colhead{$F_{\rm 0}$\tablenotemark{\tiny a}}}
\startdata
\hline
090323	&$	[0.00, 32.86]	$&$	45.1 	$&$	0.0 	$&$	14.0 	$&$	1.0 	$&$	9.8 	$&$25.16$&65.0\\
	&$	[32.86, 47.56]	$&$	42.7 	$&$	30.4 	$&$	42.0 	$&$	4.2 	$&$	1.3 	$&$16.83$&-\\
	&$	[47.56, 55.49]	$&$	20.0 	$&$	48.0 	$&$	51.0 	$&$	0.5 	$&$	1.3 	$&$12.02$&-\\
	&$	[55.49, 61.35]	$&$	37.2 	$&$	56.0 	$&$	58.0 	$&$	0.6 	$&$	1.2 	$&$7.72$&-\\
	&$	[61.35, 72.70]	$&$	42.6 	$&$	61.0 	$&$	66.0 	$&$	1.5 	$&$	9.2 	$&$6.68$&-\\
	&$	[137.18, 146.66]	$&$	56.3 	$&$	137.0 	$&$	140.4 	$&$	2.7 	$&$	4.4 	$&$3.40$&-\\
090926A	&$	[1.40, 2.80]	$&$	262.8 	$&$	0.1 	$&$	2.5 	$&$	4.8 	$&$	8.9 	$&$1.43$&98.0\\
	&$	[2.80, 3.69]	$&$	235.0 	$&$	2.9 	$&$	3.4 	$&$	0.9 	$&$	7.9 	$&$0.90$&-\\
	&$	[3.69, 5.60]	$&$	311.4 	$&$	3.5 	$&$	4.3 	$&$	1.9 	$&$	6.2 	$&$1.02$&-\\
	&$	[5.60, 9.38]	$&$	313.5 	$&$	0.0 	$&$	7.5 	$&$	1.7 	$&$	10.0 	$&$8.92$&-\\
	&$	[9.38, 9.82]	$&$	230.0 	$&$	9.3 	$&$	9.6 	$&$	5.2 	$&$	2.0 	$&$0.37$&-\\
	&$	[9.82, 10.34]	$&$	518.0 	$&$	9.8 	$&$	10.0 	$&$	1.7 	$&$	4.5 	$&$0.32$&-\\
	&$	[10.34, 12.70]	$&$	256.2 	$&$	10.2 	$&$	11.3 	$&$	1.3 	$&$	10.0 	$&$1.52$&-\\
	&$	[12.70, 14.29]	$&$	86.6 	$&$	12.9 	$&$	13.2 	$&$	2.3 	$&$	5.2 	$&$0.49$&-\\
	&$	[15.15, 16.70]	$&$	70.8 	$&$	14.9 	$&$	16.0 	$&$	5.3 	$&$	8.7 	$&$0.64$&-\\
100724B	&$	[6.00, 13.00]	$&$	55.8 	$&$	5.0 	$&$	11.0 	$&$	1.6 	$&$	7.6 	$&$7.76$&80.0\\
	&$	[13.00, 21.00]	$&$	55.0 	$&$	13.1 	$&$	18.0 	$&$	0.9 	$&$	2.3 	$&$12.01$&-\\
	&$	[37.75, 42.92]	$&$	54.8 	$&$	37.0 	$&$	38.7 	$&$	9.0 	$&$	0.9 	$&$2.78$&-\\
	&$	[42.92, 50.00]	$&$	26.7 	$&$	44.0 	$&$	46.4 	$&$	2.0 	$&$	2.8 	$&$3.40$&-\\
	&$	[53.67, 56.31]	$&$	57.8 	$&$	54.0 	$&$	55.0 	$&$	0.4 	$&$	0.9 	$&$6.06$&-\\
	&$	[56.31, 57.82]	$&$	26.6 	$&$	55.8 	$&$	56.9 	$&$	5.8 	$&$	3.9 	$&$0.80$&-\\
	&$	[57.82, 61.74]	$&$	77.4 	$&$	55.3 	$&$	60.0 	$&$	5.1 	$&$	2.4 	$&$4.27$&-\\
	&$	[61.74, 65.60]	$&$	70.8 	$&$	62.0 	$&$	63.0 	$&$	0.7 	$&$	1.1 	$&$3.84$&-\\
	&$	[65.60, 72.00]	$&$	52.0 	$&$	65.5 	$&$	67.5 	$&$	1.4 	$&$	0.8 	$&$7.36$&-\\
	&$	[72.00, 80.00]	$&$	76.0 	$&$	73.0 	$&$	75.0 	$&$	1.3 	$&$	5.6 	$&$3.06$&-\\
120226A	&$	[0.00, 16.00]	$&$	69.8 	$&$	0.7 	$&$	10.0 	$&$	1.1 	$&$	5.1 	$&$16.71$&80.0\\
	&$	[16.00, 24.00]	$&$	39.4 	$&$	14.6 	$&$	18.0 	$&$	6.8 	$&$	2.2 	$&$2.82$&-\\
	&$	[24.00, 29.00]	$&$	24.0 	$&$	24.2 	$&$	24.8 	$&$	0.1 	$&$	6.0 	$&$7.52$&-\\
	&$	[29.00, 41.50]	$&$	47.0 	$&$	29.1 	$&$	30.1 	$&$	0.1 	$&$	6.1 	$&$7.97$&-\\
	&$	[48.78, 53.79]	$&$	34.0 	$&$	48.6 	$&$	51.4 	$&$	2.0 	$&$	5.0 	$&$3.34$&-\\
	&$	[53.79, 58.15]	$&$	25.5 	$&$	53.4 	$&$	55.3 	$&$	3.0 	$&$	1.8 	$&$2.59$&-\\
130821A	&$	[23.78, 28.80]	$&$	131.0 	$&$	23.0 	$&$	27.6 	$&$	2.5 	$&$	8.8 	$&$4.25$&89.1\\
	&$	[28.80, 29.77]	$&$	50.0 	$&$	28.7 	$&$	29.4 	$&$	2.9 	$&$	9.3 	$&$0.56$&-\\
	&$	[29.77, 31.17]	$&$	140.0 	$&$	28.2 	$&$	30.6 	$&$	4.1 	$&$	4.9 	$&$1.86$&-\\
	&$	[31.17, 32.38]	$&$	129.0 	$&$	31.2 	$&$	31.7 	$&$	7.0 	$&$	0.9 	$&$0.98$&-\\
	&$	[32.38, 34.18]	$&$	198.0 	$&$	32.4 	$&$	32.9 	$&$	3.0 	$&$	1.4 	$&$0.82$&-\\
	&$	[36.09, 39.00]	$&$	120.0 	$&$	35.9 	$&$	37.3 	$&$	2.3 	$&$	6.9 	$&$1.41$&-\\
	&$	[53.75, 54.60]	$&$	90.1 	$&$	53.0 	$&$	54.3 	$&$	9.8 	$&$	6.1 	$&$0.64$&-\\
160509A	&$	[7.50, 9.50]	$&$	101.2 	$&$	6.8 	$&$	8.8 	$&$	2.9 	$&$	8.7 	$&$1.67$&100.0\\
	&$	[9.50, 12.38]	$&$	369.0 	$&$	9.0 	$&$	11.9 	$&$	2.9 	$&$	3.7 	$&$3.03$&-\\
	&$	[12.38, 13.40]	$&$	104.0 	$&$	12.5 	$&$	12.6 	$&$	0.1 	$&$	6.2 	$&$2.66$&-\\
	&$	[13.40, 14.50]	$&$	348.8 	$&$	13.5 	$&$	14.1 	$&$	1.3 	$&$	7.0 	$&$0.86$&-\\
	&$	[14.50, 16.57]	$&$	360.1 	$&$	14.6 	$&$	16.0 	$&$	0.6 	$&$	8.0 	$&$3.38$&-\\
	&$	[16.57, 20.00]	$&$	178.2 	$&$	16.2 	$&$	17.2 	$&$	3.3 	$&$	0.7 	$&$2.62$&-\\
160625B	&$	[185.51, 191.78]	$&$	1200.0 	$&$	186.9 	$&$	189.1 	$&$	2.0 	$&$	4.2 	$&$2.79$&77.0\\
	&$	[191.78, 197.09]	$&$	671.9 	$&$	189.6 	$&$	194.9 	$&$	4.1 	$&$	2.7 	$&$5.13$&-\\
	&$	[197.09, 198.56]	$&$	226.9 	$&$	195.6 	$&$	197.9 	$&$	9.9 	$&$	10.0 	$&$0.79$&-\\
	&$	[198.56, 210.00]	$&$	623.6 	$&$	196.5 	$&$	200.6 	$&$	5.4 	$&$	1.8 	$&$4.27$&-\\
170405A	&$	[5.00, 10.00]	$&$	22.1 	$&$	5.0 	$&$	6.9 	$&$	1.6 	$&$	1.0 	$&$5.13$&97.0\\
	&$	[10.00, 25.00]	$&$	41.7 	$&$	10.0 	$&$	19.4 	$&$	1.5 	$&$	1.5 	$&$20.58$&-\\
	&$	[25.00, 35.00]	$&$	60.4 	$&$	25.0 	$&$	29.3 	$&$	1.8 	$&$	9.3 	$&$5.07$&-\\
	&$	[44.90, 48.66]	$&$	62.0 	$&$	44.5 	$&$	46.9 	$&$	3.3 	$&$	2.7 	$&$2.57$&-\\
	&$	[48.66, 54.70]	$&$	45.0 	$&$	47.5 	$&$	50.6 	$&$	2.0 	$&$	6.5 	$&$3.47$&-\\
	&$	[72.00, 77.40]	$&$	29.0 	$&$	72.0 	$&$	73.7 	$&$	1.6 	$&$	2.6 	$&$2.76$&-\\
	&$	[80.21, 90.55]	$&$	21.6 	$&$	77.4 	$&$	85.0 	$&$	2.0 	$&$	8.2 	$&$8.20$&-\\
180720B	&$	[0.45, 7.49]	$&$	220.0 	$&$	0.1 	$&$	4.9 	$&$	1.8 	$&$	1.1 	$&$11.78$&79.3\\
	&$	[7.49, 10.05]	$&$	315.0 	$&$	6.9 	$&$	8.5 	$&$	6.1 	$&$	2.4 	$&$1.39$&-\\
	&$	[10.05, 13.12]	$&$	280.0 	$&$	8.6 	$&$	11.4 	$&$	3.3 	$&$	2.9 	$&$2.90$&-\\
	&$	[13.12, 13.88]	$&$	190.0 	$&$	13.1 	$&$	13.8 	$&$	0.9 	$&$	2.5 	$&$1.67$&-\\
	&$	[13.88, 14.62]	$&$	120.0 	$&$	13.9 	$&$	14.2 	$&$	2.3 	$&$	4.7 	$&$0.48$&-\\
	&$	[14.62, 15.35]	$&$	290.0 	$&$	14.6 	$&$	14.8 	$&$	2.5 	$&$	3.3 	$&$0.32$&-\\
	&$	[15.35, 17.06]	$&$	1012.0 	$&$	15.4 	$&$	16.4 	$&$	2.4 	$&$	2.2 	$&$1.53$&-\\
	&$	[17.06, 18.13]	$&$	420.0 	$&$	17.1 	$&$	17.5 	$&$	5.6 	$&$	1.8 	$&$0.61$&-\\
	&$	[18.13, 18.96]	$&$	184.0 	$&$	18.3 	$&$	18.6 	$&$	1.2 	$&$	4.3 	$&$0.58$&-\\
	&$	[18.96, 20.94]	$&$	405.0 	$&$	18.9 	$&$	19.5 	$&$	1.8 	$&$	8.9 	$&$0.63$&-\\
	&$	[21.63, 27.26]	$&$	119.4 	$&$	21.6 	$&$	23.4 	$&$	2.1 	$&$	1.9 	$&$2.85$&-\\
	&$	[27.26, 29.05]	$&$	300.0 	$&$	27.3 	$&$	28.5 	$&$	2.0 	$&$	1.8 	$&$2.19$&-\\
	&$	[29.05, 29.74]	$&$	312.0 	$&$	29.0 	$&$	29.5 	$&$	1.4 	$&$	1.5 	$&$1.15$&-\\
	&$	[29.74, 33.47]	$&$	430.0 	$&$	29.8 	$&$	29.9 	$&$	0.8 	$&$	1.6 	$&$0.37$&-\\
	&$	[47.87, 55.36]	$&$	190.0 	$&$	48.0 	$&$	50.5 	$&$	2.5 	$&$	4.1 	$&$2.69$&-\\
\hline				
\enddata
\tablenotetext{\tiny a}{In units of ${\rm counts \cdot s^{-1}}$.}
\tablenotetext{\tiny b}{In units of seconds.}
\end{deluxetable}\label{MyTabA}
\end{center}

\begin{center}
\begin{deluxetable}{lllllllllllllll}
\tablewidth{0pt} \tabletypesize{\footnotesize}
\tabletypesize{\tiny}
\tablecaption{Joint spectral fitting results of our pulses and the corresponding estimated $\Gamma$}
\tablehead{
\colhead{GRB\tablenotemark{\tiny a}}& \colhead{pulse}&
\colhead{$\alpha$}& \colhead{$\beta$}& \colhead{$E_0 ({\rm keV})$}& \colhead{$E_{\rm c}$ or $E_{\max}$\tablenotemark{\tiny b}}& \colhead{$N_{\rm 0}$ \tablenotemark{\tiny b} }& \colhead{$\chi^2_{\rm r}$ }& \colhead{$\Gamma$ \tablenotemark{\tiny c}}& \colhead{$L_{\rm \gamma,iso,52}$ \tablenotemark{\tiny d}}& & }
\startdata
\hline
090323	&$	[0.00, 32.86]	$&$	-1.19 	\pm	0.04 	$&$	-1.92 	\pm	0.16 	$&$	1281.13 	\pm	422.49 	$&$	22.27 	\pm	7.34 	$&$	2.69 	\pm	0.32 	$&$	1.36 	$&$	185.59 	\pm	66.07 	$&$	7.44 	\pm	0.95 	$	\\
	&$	[32.86, 47.56]	$&$	-0.74 	\pm	0.08 	$&$	-2.12 	\pm	0.31 	$&$	185.00 	\pm	40.17 	$&$	19.21 	\pm	18.69 	$&$	0.75 	\pm	0.20 	$&$	1.17 	$&$	142.13 	\pm	140.10 	$&$	4.37 	\pm	0.82 	$	\\
	&$	[47.56, 55.49]	$&$	-1.06 	\pm	0.05 	$&$	-3.28 	\pm	0.55 	$&$	844.73 	\pm	257.40 	$&$	105.27 			$&$	2.48 	\pm	0.61 	$&$	0.86 	$&$	\gtrsim165.80 	\pm	164.10 	$&$	6.72 	\pm	1.53 	$	\\
	&$	[55.49, 61.35]	$&$	-0.93 	\pm	0.05 	$&$	-2.67 	\pm	0.10 	$&$	710.47 	\pm	172.87 	$&$	319.53 			$&$	1.90 	\pm	0.47 	$&$	1.16 	$&$	\gtrsim273.47 	\pm	33.97 	$&$	11.16 	\pm	1.50 	$	\\
	&$	[61.35, 72.70]	$&$	-0.97 	\pm	0.04 	$&$	-2.12 	\pm	0.06 	$&$	454.05 	\pm	82.64 	$&$	64.79 	\pm	16.43 	$&$	2.41 	\pm	0.35 	$&$	1.08 	$&$	265.42	\pm	30.72 	$&$	11.01 	\pm	1.02 	$	\\
	&$	[137.18, 146.66]	$&$	-0.94 	\pm	0.19 	$&$	-1.93 	\pm	0.13 	$&$	100.13 	\pm	44.35 	$&$	40.00 			$&$	3.14 	\pm	3.75 	$&$	1.16 	$&$	\gtrsim272.18 	\pm	79.58 	$&$	11.95 	\pm	10.58 	$	\\
090926A	&$	[1.40, 2.80]	$&$	-0.49 	\pm	0.04 	$&$	-2.75 	\pm	0.07 	$&$	261.24 	\pm	16.54 	$&$	130.56 			$&$	1.72 	\pm	0.32 	$&$	0.94 	$&$ \gtrsim223.16	\pm	21.44 	$&$	15.68 	\pm	0.83 	$	\\	
	&$	[2.80, 3.69]	$&$	-0.52 	\pm	0.03 	$&$	-3.16 	\pm	0.17 	$&$	286.94 	\pm	14.79 	$&$	175.46 			$&$	3.18 	\pm	0.47 	$&$	0.98 	$&$	\gtrsim233.04 	\pm	50.79 	$&$	25.34 	\pm	1.27 	$	\\
	&$	[3.69, 5.60]	$&$	-0.59 	\pm	0.03 	$&$	-2.32 	\pm	0.03 	$&$	232.59 	\pm	11.04 	$&$	200.31 	\pm	62.39 	$&$	5.06 	\pm	0.46 	$&$	1.01 	$&$	379.60	\pm	33.24 	$&$	32.22 	\pm	1.23 	$	\\
	&$	[5.60, 9.38]	$&$	-0.71 	\pm	0.02 	$&$	-2.31 	\pm	0.01 	$&$	254.13 	\pm	11.12 	$&$	1216.72 			$&$	5.89 	\pm	0.60 	$&$	1.31 	$&$	\gtrsim294.12 	\pm	4.72	$&$	23.92 	\pm	0.75 	$	\\
	&$	[9.38, 9.82]	$&$	-0.51 	\pm	0.05 	$&$	-2.55 	\pm	0.05 	$&$	215.07 	\pm	18.93 	$&$	3321.62 			$&$	3.50 	\pm	0.92 	$&$	0.94 	$&$	\gtrsim639.87 	\pm	41.22 	$&$	24.50 	\pm	13.04 	$	\\
	&$	[9.82, 10.34]	$&$	-1.14 	\pm	0.05 	$&$	-1.96 	\pm	0.02 	$&$	271.16 	\pm	50.56 	$&$	702.56 	\pm	269.31 	$&$	55.76 	\pm	8.84 	$&$	0.96 	$&$	682.42	\pm	59.56 	$&$	48.01 	\pm	5.63 	$	\\
	&$	[10.34, 12.70]	$&$	-0.86 	\pm	0.03 	$&$	-2.22 	\pm	0.01 	$&$	197.77 	\pm	12.59 	$&$	2040.68 			$&$	14.02 	\pm	1.97 	$&$	1.35 	$&$	\gtrsim510.09 	\pm	9.74 	$&$	22.47 	\pm	10.88 	$	\\
	&$	[12.70, 14.29]	$&$	-1.01 	\pm	0.06 	$&$	-2.15 	\pm	0.02 	$&$	201.47 	\pm	34.37 	$&$	935.76 			$&$	11.15 	\pm	3.62 	$&$	1.01 	$&$	\gtrsim524.87 	\pm	21.14 	$&$	9.92 	\pm	4.37 	$	\\
	&$	[15.15, 16.70]	$&$	-1.07 	\pm	0.10 	$&$	-2.11 	\pm	0.02 	$&$	171.16 	\pm	47.51 	$&$	2662.02 			$&$	10.44 	\pm	5.68 	$&$	0.96 	$&$	\gtrsim539.92 	\pm	25.69 	$&$	6.49 	\pm	5.99 	$	\\
100724B	&$	[6.00, 13.00]	$&$	-0.69 	\pm	0.05 	$&$	-1.75 	\pm	0.05 	$&$	410.51 	\pm	57.30 	$&$	20.03 	\pm	4.85 	$&$	0.77 	\pm	0.13 	$&$	1.11 	$&$	78.39	\pm	18.99 	$&$	2.45 	\pm	0.25 	$	\\
	&$	[13.00, 21.00]	$&$	-0.75 	\pm	0.03 	$&$	-1.76 	\pm	0.03 	$&$	440.29 	\pm	47.42 	$&$	27.06 	\pm	2.60 	$&$	1.29 	\pm	0.16 	$&$	1.16 	$&$	105.91	\pm	10.16 	$&$	3.42 	\pm	0.18 	$	\\
	&$	[37.75, 42.92]	$&$	-0.87 	\pm	0.05 	$&$	-2.16 	\pm	0.13 	$&$	343.12 	\pm	51.44 	$&$	42.09 	\pm	23.75 	$&$	1.76 	\pm	0.32 	$&$	1.09 	$&$	125.49	\pm	34.97 	$&$	1.05 	\pm	0.14 	$	\\
	&$	[42.92, 50.00]	$&$	-0.85 	\pm	0.05 	$&$	-2.31 	\pm	0.13 	$&$	272.64 	\pm	36.59 	$&$	69.97 	\pm	61.06 	$&$	1.54 	\pm	0.27 	$&$	1.41 	$&$	116.18	\pm	32.37 	$&$	1.05 	\pm	0.14 	$	\\
	&$	[53.67, 56.31]	$&$	-0.83 	\pm	0.05 	$&$	-2.90 	\pm	0.20 	$&$	426.70 	\pm	51.85 	$&$	70.00 			$&$	1.95 	\pm	0.44 	$&$	0.94 	$&$	\gtrsim75.11 	\pm	20.53 	$&$	1.14 	\pm	0.15 	$	\\
	&$	[56.31, 57.82]	$&$	-0.86 	\pm	0.05 	$&$	-2.99 	\pm	0.28 	$&$	507.02 	\pm	75.44 	$&$	104.77 			$&$	2.41 	\pm	0.61 	$&$	0.80 	$&$	\gtrsim130.72 	\pm	58.87 	$&$	1.40 	\pm	0.20 	$	\\
	&$	[57.82, 61.74]	$&$	-0.75 	\pm	0.03 	$&$	-2.51 	\pm	0.11 	$&$	353.25 	\pm	26.69 	$&$	94.98 	\pm	77.81 	$&$	2.16 	\pm	0.25 	$&$	0.96 	$&$	114.41	\pm	26.65 	$&$	1.95 	\pm	0.13 	$	\\
	&$	[61.74, 65.60]	$&$	-0.74 	\pm	0.03 	$&$	-2.01 	\pm	0.06 	$&$	377.22 	\pm	34.16 	$&$	28.44 	\pm	5.73 	$&$	2.27 	\pm	0.27 	$&$	1.13 	$&$	111.32	\pm	22.44 	$&$	3.46 	\pm	0.22 	$	\\
	&$	[65.60, 72.00]	$&$	-0.82 	\pm	0.03 	$&$	-2.35 	\pm	0.10 	$&$	422.82 	\pm	29.82 	$&$	41.19 	\pm	18.96 	$&$	2.58 	\pm	0.24 	$&$	1.11 	$&$	96.73 	\pm	19.63 	$&$	2.06 	\pm	0.13 	$	\\
	&$	[72.00, 80.00]	$&$	-0.77 	\pm	0.03 	$&$	-2.23 	\pm	0.06 	$&$	349.92 	\pm	24.66 	$&$	52.16 	\pm	15.58 	$&$	1.96 	\pm	0.19 	$&$	1.37 	$&$	142.98	\pm	18.25 	$&$	1.91 	\pm	0.10 	$	\\
120226A	&$	[0.00, 16.00]	$&$	-0.82 	\pm	0.04 	$&$	-2.53 	\pm	0.03 	$&$	290.64 	\pm	25.47 	$&$	9603.43 			$&$	1.22 	\pm	0.22 	$&$	1.36 	$&$	\gtrsim211.35 	\pm	7.02 	$&$	0.62 	\pm	0.04 	$	\\
	&$	[16.00, 24.00]	$&$	-0.89 	\pm	0.06 	$&$	-2.10 	\pm	0.17 	$&$	281.99 	\pm	46.95 	$&$	20.22 	\pm	9.80 	$&$	1.82 	\pm	0.34 	$&$	1.10 	$&$	79.13	\pm	38.36 	$&$	0.76 	\pm	0.15 	$	\\
	&$	[24.00, 29.00]	$&$	-0.58 	\pm	0.15 	$&$	-2.10 	\pm	0.17 	$&$	122.68 	\pm	32.22 	$&$	29.76 	\pm	21.52 	$&$	0.52 	\pm	0.24 	$&$	1.06 	$&$	75.94	\pm	28.15 	$&$	0.45 	\pm	0.14 	$	\\
	&$	[29.00, 41.50]	$&$	-0.92 	\pm	0.06 	$&$	-2.10 	\pm	0.12 	$&$	259.07 	\pm	44.36 	$&$	40.00 			$&$	1.72 	\pm	0.49 	$&$	1.14 	$&$	\gtrsim95.37 	\pm	21.99 	$&$	1.02 	\pm	0.12 	$	\\
	&$	[48.78, 53.79]	$&$	-0.78 	\pm	0.20 	$&$	-2.08 	\pm	0.24 	$&$	163.34 	\pm	62.49 	$&$	40.00 			$&$	0.97 	\pm	1.18 	$&$	0.95 	$&$	\gtrsim98.41	\pm	61.70 	$&$	0.52 	\pm	0.36 	$	\\
	&$	[53.79, 58.15]	$&$	-1.21 	\pm	0.15 	$&$	-2.21 	\pm	0.61 	$&$	301.73 	\pm	191.57 	$&$	40.00 			$&$	3.72 	\pm	3.17 	$&$	1.27 	$&$	\gtrsim91.42 	\pm	9.10 	$&$	0.30 	\pm	0.26 	$	\\
130821A	&$	[23.78, 28.80]	$&$	-0.81 	\pm	0.05 	$&$	-2.19 	\pm	0.14 	$&$	231.00 	\pm	30.63 	$&$	20.00 	\pm	9.34 	$&$	2.20 	\pm	0.40 	$&$	1.00 	$&$	78.29	\pm	36.56 	$&$	1.00 	\pm	0.13 	$	\\
	&$	[28.80, 29.77]	$&$	-0.81 	\pm	0.05 	$&$	-2.19 	\pm	0.14 	$&$	231.00 	\pm	30.63 	$&$	20.00 	\pm	9.34 	$&$	2.20 	\pm	0.40 	$&$	1.00 	$&$	78.31	\pm	24.71 	$&$	1.00 	\pm	0.13 	$	\\
	&$	[29.77, 31.17]	$&$	-0.88 	\pm	0.05 	$&$	-2.09 	\pm	0.11 	$&$	319.77 	\pm	44.37 	$&$	20.01 	\pm	6.31 	$&$	4.54 	\pm	0.74 	$&$	0.99 	$&$	78.31	\pm	24.71 	$&$	2.25 	\pm	0.27 	$	\\
	&$	[31.17, 32.38]	$&$	-0.79 	\pm	0.06 	$&$	-2.86 	\pm	0.26 	$&$	219.54 	\pm	25.54 	$&$	35.09 			$&$	4.12 	\pm	1.19 	$&$	0.90 	$&$	\gtrsim89.13 	\pm	38.59 	$&$	1.36 	\pm	0.23 	$	\\
	&$	[32.38, 34.18]	$&$	-1.00 	\pm	0.04 	$&$	-2.42 	\pm	0.24 	$&$	356.65 	\pm	49.06 	$&$	29.93 	\pm	24.13 	$&$	7.63 	\pm	1.18 	$&$	1.06 	$&$	117.15	\pm	94.45 	$&$	1.77 	\pm	0.21 	$	\\
	&$	[36.09, 39.00]	$&$	-0.99 	\pm	0.07 	$&$	-2.10 	\pm	0.11 	$&$	229.52 	\pm	46.97 	$&$	40.00 			$&$	5.51 	\pm	1.92 	$&$	1.26 	$&$	\gtrsim145.57 	\pm	33.29 	$&$	1.94 	\pm	0.23 	$	\\
	&$	[53.75, 54.60]	$&$	-1.12 	\pm	0.22 	$&$	-2.09 	\pm	0.39 	$&$	259.94 	\pm	184.96 	$&$	40.00 			$&$	7.76 	\pm	11.76 	$&$	1.12 	$&$	\gtrsim138.43 	\pm	130.10 	$&$	0.95 	\pm	0.59 	$	\\
160509A	&$	[7.50, 9.50]	$&$	-0.81 	\pm	0.06 	$&$	-2.73 	\pm	0.09 	$&$	590.75 	\pm	94.35 	$&$	168.35 			$&$	1.43 	\pm	0.40 	$&$	0.85 	$&$	\gtrsim154.16 	\pm	19.22 	$&$	1.97 	\pm	0.17 	$	\\
	&$	[9.50, 12.38]	$&$	-0.72 	\pm	0.03 	$&$	-2.19 	\pm	0.03 	$&$	432.30 	\pm	26.93 	$&$	65.21 	\pm	8.84 	$&$	3.33 	\pm	0.32 	$&$	1.06 	$&$	173.26	\pm	10.40 	$&$	8.12 	\pm	0.28 	$	\\
	&$	[12.38, 13.40]	$&$	-0.72 	\pm	0.04 	$&$	-2.15 	\pm	0.04 	$&$	343.81 	\pm	29.63 	$&$	76.19 	\pm	17.52 	$&$	5.33 	\pm	0.71 	$&$	0.96 	$&$	167.00	\pm	14.03 	$&$	11.00 	\pm	0.57 	$	\\
	&$	[13.40, 14.50]	$&$	-0.66 	\pm	0.03 	$&$	-2.25 	\pm	0.05 	$&$	281.84 	\pm	19.20 	$&$	47.58 	\pm	11.34 	$&$	5.71 	\pm	0.67 	$&$	0.80 	$&$	202.07	\pm	48.14 	$&$	10.83 	\pm	0.52 	$	\\
	&$	[14.50, 16.57]	$&$	-0.78 	\pm	0.02 	$&$	-2.35 	\pm	0.04 	$&$	288.36 	\pm	14.73 	$&$	84.73 	\pm	22.84 	$&$	9.33 	\pm	0.75 	$&$	0.99 	$&$	150.96	\pm	12.74 	$&$	8.79 	\pm	0.29 	$	\\
	&$	[16.57, 20.00]	$&$	-0.77 	\pm	0.02 	$&$	-2.00 	\pm	0.01 	$&$	275.95 	\pm	15.51 	$&$	82.78 	\pm	6.22 	$&$	7.56 	\pm	0.59 	$&$	1.09 	$&$	225.32	\pm	7.16 	$&$	11.52 	\pm	0.27 	$	\\
160625B	&$	[185.51, 191.78]	$&$	-0.66 	\pm	0.01 	$&$	-2.25 	\pm	0.01 	$&$	682.14 	\pm	16.09 	$&$	43.87 	\pm	2.02 	$&$	4.87 	\pm	0.18 	$&$	2.43 	$&$	206.56	\pm	9.51 	$&$	37.00 	\pm	0.37 	$	\\
	&$	[191.78, 197.09]	$&$	-0.67 	\pm	0.01 	$&$	-2.82 	\pm	0.01 	$&$	435.58 	\pm	8.76 	$&$	2766.50 			$&$	6.70 	\pm	0.30 	$&$	1.88 	$&$	\gtrsim354.14 	\pm	5.48 	$&$	21.40 	\pm	0.27 	$	\\
	&$	[197.09, 198.56]	$&$	-0.69 	\pm	0.02 	$&$	-2.74 	\pm	0.05 	$&$	506.33 	\pm	21.84 	$&$	139.69 	\pm	72.12 	$&$	6.31 	\pm	0.47 	$&$	0.88 	$&$	253.08 	\pm	39.15 	$&$	21.58 	\pm	0.76 	$	\\
	&$	[198.56, 210.00]	$&$	-0.71 	\pm	0.01 	$&$	-2.59 	\pm	0.01 	$&$	487.43 	\pm	9.34 	$&$	213.08 	\pm	27.42 	$&$	5.39 	\pm	0.17 	$&$	2.18 	$&$	238.91	\pm	8.18 	$&$	17.10 	\pm	0.17 	$	\\
170405A	&$	[5.00, 10.00]	$&$	-0.81 	\pm	0.16 	$&$	-2.07 	\pm	0.25 	$&$	323.74 	\pm	126.13 	$&$	40.00 			$&$	0.71 	\pm	0.60 	$&$	1.12 	$&$	\gtrsim204.86 	\pm	134.02 	$&$	8.32 	\pm	2.85 	$	\\
	&$	[10.00, 25.00]	$&$	-0.61 	\pm	0.06 	$&$	-2.11 	\pm	0.06 	$&$	222.88 	\pm	24.96 	$&$	44.81 	\pm	13.77 	$&$	0.50 	\pm	0.11 	$&$	1.22 	$&$	170.37	\pm	20.77 	$&$	7.59 	\pm	0.78 	$	\\
	&$	[25.00, 35.00]	$&$	-0.74 	\pm	0.05 	$&$	-2.23 	\pm	0.07 	$&$	257.01 	\pm	27.19 	$&$	58.83 	\pm	25.48 	$&$	1.18 	\pm	0.21 	$&$	1.13 	$&$	251.51	\pm	38.43 	$&$	8.88 	\pm	0.83 	$	\\
	&$	[44.90, 48.66]	$&$	-0.86 	\pm	0.08 	$&$	-2.10 	\pm	0.15 	$&$	295.34 	\pm	66.05 	$&$	40.00 			$&$	1.89 	\pm	0.79 	$&$	1.01 	$&$	\gtrsim269.56 	\pm	82.11 	$&$	15.93 	\pm	2.52 	$	\\
	&$	[48.66, 54.70]	$&$	-0.93 	\pm	0.08 	$&$	-2.10 	\pm	0.13 	$&$	255.39 	\pm	53.20 	$&$	40.00 			$&$	2.37 	\pm	0.90 	$&$	1.08 	$&$	\gtrsim251.31 	\pm	66.73 	$&$	12.12 	\pm	1.64 	$	\\
	&$	[72.00, 77.40]	$&$	-0.85 	\pm	0.13 	$&$	-2.17 	\pm	0.29 	$&$	266.62 	\pm	88.83 	$&$	40.00 			$&$	1.04 	\pm	0.73 	$&$	1.37 	$&$	\gtrsim232.18 	\pm	180.10 	$&$	5.61 	\pm	1.92 	$	\\
	&$	[80.21, 90.55]	$&$	-1.00 	\pm	0.09 	$&$	-2.34 	\pm	0.39 	$&$	317.66 	\pm	84.38 	$&$	40.00 			$&$	1.57 	\pm	0.69 	$&$	1.37 	$&$	\gtrsim160.30 	\pm	150.10 	$&$	3.44 	\pm	0.94 	$	\\
180720B	&$	[0.45, 7.49]	$&$	-1.03 	\pm	0.01 	$&$	-2.71 	\pm	0.29 	$&$	1333.44 	\pm	96.76 	$&$	20.07 	\pm	12.86 	$&$	6.73 	\pm	0.37 	$&$	1.04 	$&$	60.49	\pm	37.66 	$&$	1.74 	\pm	0.23 	$	\\
	&$	[7.49, 10.05]	$&$	-1.03 	\pm	0.02 	$&$	-2.35 	\pm	0.14 	$&$	1070.04 	\pm	81.10 	$&$	20.00 	\pm	6.98 	$&$	14.57 	\pm	0.87 	$&$	1.13 	$&$	64.74	\pm	22.60 	$&$	3.40 	\pm	0.78 	$	\\
	&$	[10.05, 13.12]	$&$	-0.98 	\pm	0.01 	$&$	-2.31 	\pm	0.07 	$&$	789.44 	\pm	47.48 	$&$	21.26 	\pm	3.62 	$&$	13.57 	\pm	0.75 	$&$	0.98 	$&$	68.80 	\pm	11.73 	$&$	3.38 	\pm	0.12 	$	\\
	&$	[13.12, 13.88]	$&$	-0.94 	\pm	0.04 	$&$	-2.76 	\pm	0.10 	$&$	452.06 	\pm	47.25 	$&$	114.60 			$&$	11.80 	\pm	2.01 	$&$	0.77 	$&$	\gtrsim99.81 	\pm	13.72 	$&$	1.92 	\pm	0.14 	$	\\
	&$	[13.88, 14.62]	$&$	-1.05 	\pm	0.03 	$&$	-2.94 	\pm	0.18 	$&$	643.98 	\pm	68.59 	$&$	54.01 			$&$	20.83 	\pm	2.85 	$&$	1.05 	$&$	\gtrsim112.54	\pm	31.82 	$&$	2.36 	\pm	0.18 	$	\\
	&$	[14.62, 15.35]	$&$	-0.99 	\pm	0.03 	$&$	-2.71 	\pm	0.09 	$&$	573.22 	\pm	57.59 	$&$	73.11 			$&$	17.07 	\pm	2.43 	$&$	0.89 	$&$	\gtrsim150.61 	\pm	21.95 	$&$	2.64 	\pm	0.18 	$	\\
	&$	[15.35, 17.06]	$&$	-0.88 	\pm	0.01 	$&$	-2.00 	\pm	0.04 	$&$	651.68 	\pm	36.34 	$&$	20.00 	\pm	4.88 	$&$	18.06 	\pm	1.01 	$&$	1.02 	$&$	64.74	\pm	15.81 	$&$	8.88 	\pm	0.44 	$	\\
	&$	[17.06, 18.13]	$&$	-0.98 	\pm	0.02 	$&$	-2.49 	\pm	0.12 	$&$	563.70 	\pm	35.96 	$&$	20.01 	\pm	11.18 	$&$	26.78 	\pm	1.84 	$&$	1.12 	$&$	64.75	\pm	36.17 	$&$	4.68 	\pm	0.25 	$	\\
	&$	[18.13, 18.96]	$&$	-1.10 	\pm	0.03 	$&$	-2.83 	\pm	0.15 	$&$	380.05 	\pm	35.80 	$&$	60.74 			$&$	33.60 	\pm	4.95 	$&$	0.80 	$&$	\gtrsim106.19 	\pm	24.73 	$&$	1.78 	\pm	0.14 	$	\\
	&$	[18.96, 20.94]	$&$	-1.32 	\pm	0.02 	$&$	-2.75 	\pm	0.08 	$&$	670.60 	\pm	76.43 	$&$	279.93 			$&$	48.05 	\pm	5.18 	$&$	0.99 	$&$	\gtrsim169.11 	\pm	18.84 	$&$	1.23 	\pm	0.06 	$	\\
	&$	[21.63, 27.26]	$&$	-1.40 	\pm	0.02 	$&$	-2.56 	\pm	0.06 	$&$	517.68 	\pm	64.80 	$&$	156.22 			$&$	35.01 	\pm	4.14 	$&$	1.02 	$&$	\gtrsim104.25 	\pm	8.45 	$&$	0.53 	\pm	0.03 	$	\\
	&$	[27.26, 29.05]	$&$	-1.26 	\pm	0.05 	$&$	-2.45 	\pm	0.19 	$&$	334.43 	\pm	59.97 	$&$	125.89 	\pm	224.09 	$&$	23.44 	\pm	3.41 	$&$	0.74 	$&$	\gtrsim96.05 	\pm	36.97 	$&$	0.57 	\pm	0.07 	$	\\
	&$	[29.05, 29.74]	$&$	-0.98 	\pm	0.03 	$&$	-2.79 	\pm	0.12 	$&$	493.63 	\pm	50.78 	$&$	84.16 			$&$	16.36 	\pm	2.54 	$&$	0.90 	$&$	\gtrsim102.41 	\pm	17.76 	$&$	2.28 	\pm	0.17 	$	\\
	&$	[29.74, 33.47]	$&$	-1.23 	\pm	0.02 	$&$	-2.91 	\pm	0.09 	$&$	795.84 	\pm	81.94 	$&$	398.48 			$&$	20.48 	\pm	1.97 	$&$	0.88 	$&$	\gtrsim217.42 	\pm	27.22 	$&$	0.99 	\pm	0.05 	$	\\
	&$	[47.87, 55.36]	$&$	-1.23 	\pm	0.03 	$&$	-2.59 	\pm	0.08 	$&$	564.71 	\pm	69.22 	$&$	398.45 			$&$	11.15 	\pm	1.42 	$&$	1.07 	$&$	\gtrsim135.43 	\pm	14.21 	$&$	0.46 	\pm	0.03 	$	\\
\hline				
\enddata
\tablenotetext{\tiny a}{Redshift measurements reference. GRB~090323, $z$=3.57,
\cite{Ackermann_M-2013-Ajello_M-ApJS.209.11A}. GRB~090926A, $z$=2.1062,
\cite{Ackermann_M-2013-Ajello_M-ApJS.209.11A}. GRB~160509A, $z$=1.17,
\cite{Tanvir_NR-2016-Levan_AJ-GCN.19419.1T}. GRB~160625B, $z$=1.406,
\cite{D'Elia_V-2016-Melandri_A-GCN.19601.1D}. GRB~170405A, $z$=3.51,
\cite{deUgartePostigo_A-2017-Kann_DA-GCN.20990.1D}. GRB~180720B, $z$=0.654,
\cite{Vreeswijk_PM-2018-Kann_DA-GCN.22996.1V}. Bursts with no redshift measurements are assumed to have $z=1.0$.}
\tablenotetext{\tiny b}{$E_{\rm c}$ and $E_{\max}$ are in the unit of MeV and $N_0$ is in unit of ${\rm photons \cdot cm^{-2} \cdot s^{-1}\cdot keV^{-1}}$.}
\tablenotetext{c}{The exact value of $\Gamma$ is derived from Equation~(\ref{Eq_Gamma})
and the lower limit of $\Gamma$ is from $\Gamma>\Gamma_{\uparrow}$ based on Equation~(\ref{Eq_Gamma_low}).}
\tablenotetext{d}{$L_{\rm \gamma,iso,52}$ is estimated in the 8~keV-100~GeV energy band.}
\end{deluxetable}\label{MyTabB}
\end{center}

\begin{center}
\begin{deluxetable}{ccc|ccc}
\tablewidth{0pt} \tabletypesize{\footnotesize}
\tablecaption{Initial Lorentz factor of the jet producing the external-shock afterglow}
\tablehead{
\colhead{GRB}& \colhead{$\Gamma_{0,\rm ps}$} & \colhead{Pulses for estimating $\Gamma_{0,\rm ps}$}&
\colhead{$\Gamma_{0,\rm es}$} & \colhead{References for $\Gamma_{0,\rm es}$} }
\startdata
\hline
090323	& $\gtrsim 206.92$&  0-150~s&$	350-870	$ (ISM) &\cite{Ackermann_M-2013-Ajello_M-ApJS.209.11A}\\
 &$\gtrsim 183.45	$& $\lesssim 70$~s &$	350-590	$ (wind)& \\
\hline
090926A & $\gtrsim 357.24$&  0-18~s&$	750	$ for ISM (lower limit)&\cite{Ackermann_M-2011-Ajello_M-ApJ.729.114A}\\
 &$\gtrsim 304.08	$&  $\lesssim 10$~s &$	290	$ for wind (lower limit)& \\
  && &$	520-700	$ for ISM&\cite{Ackermann_M-2013-Ajello_M-ApJS.209.11A}\\
  & & &$	400-450	$ for wind& \\
\hline
100724B & $\gtrsim 104.58$ & 0-80~s& &  \\
\hline
120226A	& $\gtrsim 106.49$ & 0-60~s& &  \\
\hline
130821A & $\gtrsim 98.67$ & 20-60~s&$	440	$&\cite{Liang_YF-2014-Zhou_B-ApJ.781.74L}\\
\hline
160509A & $\gtrsim 185.67$ & 5-20~s &$	600	$ &\cite{Fraija_N-2020-Laskar_T-ApJ.905.112F}\\
\hline
160625B	& $\gtrsim 239.92$ &  185-210~s &Reverse Shock (Model 1)(ISM) $	290	$ &\cite{Alexander_KD-2017-Laskar_T-ApJ.848.69A}\\
 && &Reverse Shock (Model 2)(ISM) $	120	$ &\\
 && &Reverse Shock (Model 3)(ISM) $	370	$ &\\
 && &$	500	$ (wind-to-ISM) &\cite{Fraija_N-2017-Veres_P-ApJ.848.15F}\\
\hline
170405A & $\gtrsim 212.04$ &  0-100~s &$	420	$ (ISM) (upper limit)&\cite{Arimoto_M-2020-Asano_K-ApJ.891.106A}\\
 && & or wind medium depend on specific parameters& \\
\hline
180720B & $\gtrsim 75.71$ & 0-60~s &$142(80)$ for wind and $v_{\rm w}=10^3(10^2)$\,km s$^{-1}$ & \cite{Ronchi_M-2020-Fumagalli_F-AA.636A.55R}\\
 && &$294 (220)$ for ISM and $n_0=1(10)$\,cm$^{-3}$& \\
\hline				
\enddata
\end{deluxetable}\label{MyTabC}
\end{center}


\clearpage

\end{document}